\date{}
\documentclass[preprint,prd,12pt]{revtex4-1}
\usepackage{amsmath}
\usepackage{amssymb}
\usepackage{mathrsfs} 
\usepackage{graphicx} \usepackage{bm} \usepackage{epsfig}
\usepackage{epstopdf} 
\DeclareGraphicsExtensions{.pdf,.eps,.png,.jpg,.mps} 
\usepackage{color}
\usepackage[centerlast]{caption}
\usepackage{subcaption}
\usepackage[english]{babel}
\def\Tr{\mathop{\rm Tr}}

\begin{document}

\title{The Hartree approximation in curved spacetimes revisited I: the effective potential in de Sitter}

\author{Diana L. L\'opez Nacir $^{1,2}$}
\author{Francisco D. Mazzitelli$^{3}$}
\author{Leonardo G. Trombetta$^2$}

\affiliation{$^1$ Abdus Salam International Centre for Theoretical Physics
Strada Costiera 11, 34151, Trieste, Italy}
\affiliation{$^2$ Departamento de F\'\i sica and IFIBA, FCEyN UBA, Facultad de Ciencias Exactas y Naturales,
 Ciudad Universitaria, Pabell\' on I, 1428 Buenos Aires, Argentina}
\affiliation{$^3$ Centro At\'omico Bariloche 
Comisi\'on Nacional de Energ\'\i a At\'omica,
R8402AGP Bariloche, Argentina}

\date{\today}

\begin{abstract}
We consider a quantum scalar field with $\lambda\phi^4$ interaction in curved spacetimes. The quantum effects are taken into account nonperturbatively using
the Hartree approximation to the 2PI effective action.  Although this approximation has been considered in many previous works, we reconsider it using a
consistent nonperturbative renormalization procedure, which we extend to general curved spacetimes. We obtain the renormalized equations for the mean field and for the propagator of the fluctuations, showing explicitly their independence on the arbitrary scale introduced by the regularization scheme. We apply our results to the particular case of de Sitter spacetime and discuss spontaneous symmetry breaking. The results  depend strongly on the renormalization procedure.

\end{abstract}


\maketitle

\section{Introduction}

In the last years there has been a renewed interest in the analysis of interacting field theories in de Sitter spacetimes. On the one hand, 
there is observational evidence that supports the existence of periods of exponential or quasi-exponential expansion of the universe, 
both during the inflationary era  before the radiation dominated period in the early universe, 
and in the present accelerated expansion. Quantum fields may play a relevant role during both periods.  On the other hand, the high degree of
symmetry of de Sitter spacetime, make it a very interesting arena to analyze conceptual problems of semiclassical and quantum gravity.

There are several unsettled questions regarding interacting quantum fields in de Sitter spacetime.  In particular, perturbative calculations 
show the existence of secular growth and/or infrared (IR) divergences \cite{Weinberg,Woodard,Seery1,SeeryR,TanakaR,Meulen,Shandera}. 
The appearance of terms that secularly grow with time, which can lead to a breakdown of perturbation theory at late times, might be due 
to a deficiency of the perturbative approach. Therefore, it is clear that in order to make reliable predictions, such infrared effects would
have to be understood, and for this the use of nonperturbative techniques seems to be unavoidable. There are several works in the literature
attempting to understand the physical effects of the secular growth  \cite{Meulen,Starobinski,Shandera,Woodard,Rajaraman,Hollands,Beneke,Seery1,SeeryR,TanakaR,Rigopoulos}.

Recently, there has been much progress in the context of single-field inflationary models where there are only adiabatic perturbations, and the 
variable $\zeta$ that describes the curvature perturbations remains constant on super-horizon scales  
\cite{Matias1,Matias3,MatiasL,Baumann,Tanaka,TanakaR,Woodardult}. Working directly with the evolution operator for $\zeta$,  it has been shown  
\cite{Matias3,MatiasL,Tanaka,TanakaR,Baumann} that despite the existence of individual contributions to $\zeta$ correlation functions that secularly 
grow with time, when all of them are added up cancellations occur so that the final result does not exhibit a secular growth.
For this, the use of symmetries and nonperturbative techniques have shown to be crucial and the computations to be very subtle, 
even for the case in which only  scalar perturbations are considered. A complete explicit calculation that addresses the renormalization
procedure is extremely complex, and still lacking. An alternative two-step procedure to perturbatively compute loop corrections to $\zeta$ 
correlation functions has been also applied (see for instance \cite{Bartolo,Seery1}). The first step consists in calculating correlation functions 
of the inflaton fluctuations, and the second in obtaining the ones of $\zeta$ by performing a gauge transformation \cite{Meulen,Bartolo,Seery1,SeeryR}. In this case, even though the correlation functions of the inflaton fluctuations can show secular growth, the use of nonperturbative schemes is avoided by performing the gauge transformation soon after the field modes exited the horizon (which would be well before the secular growing effect becomes 
relevant). Nevertheless, to our knowledge, when loop corrections are included there is still no formal general proof of the equivalence between the two procedures.

The effect of secular growth is less clear for massless interacting test fields in de Sitter space (which in the context of inflationary models 
could be interpreted as an approximation to entropy modes). This kind of fields have been considered by several authors in different frameworks;
for instance, in the stochastic approach \cite{Starobinski,Woodard,ProkopecN}, in the context of Euclidean de Sitter space  \cite{Rajaraman,Hollands,Beneke},
using  dynamical renormalization group resummations \cite{Shandera},  or  other nonperturbative approximation schemes  \cite{Riotto,Boyanovsky,Akhmedov, Rigopoulos, Serreau,Youssef,Arai,Serreau0}.    In particular, it has been pointed out that  in the presence of interactions a scalar field  dynamically acquires a mass \cite{Starobinski}, which screens perturbative IR divergences. The dynamical  generated mass has been obtained in the different approaches and, although the numerical value does not coincide for some of the approximations,  they agree on that it is non-analytic 
in the coupling constant.  Also, it has been argued that for scalar fields presenting  a symmetry breaking  tree-level potential,
the presence of large infrared effects could give rise to a symmetry restoration \cite{Ratra} (see also \cite{Arai, Serreau0,ProkopecN}). 
Nevertheless, the conditions under which the symmetry is restored and the order of the phase transition are still unclear \cite{Arai,Serreau0,ProkopecN}. 
We will come back to these  points later. 

Nonperturbative  quantum field theory is a difficult topic even in flat spacetime. The most widely used approximations 
involve the selective summation of Feynman diagrams of infinite perturbative order, like the Hartree approximation, or 
expansions in $1/N$, where $N$ is the number of quantum fields (or the number of colors in QCD). Nonperturbative calculations can 
be addressed using the effective action for composite operators introduced a long time ago by Cornwall, Jackiw and Tomboulis \cite{CJT}, 
also called in the literature two-particle irreducible effective action (2PI EA). Indeed, this is a general and systematically improvable 
approach for resumming classes of Feynman diagrams. In its lowest approximation level, it corresponds to the so called Hartree (or Gaussian) approximation. 

The Gaussian approximation can also be introduced without referring to the 2PI EA, by means of a variational principle. It has been the subject of many works
 after the papers by Stevenson \cite{Stevenson}, in which the author showed that it is possible to  choose carefully the bare constants of the $\lambda\phi^4$
 theory allowing for the emergence of a nontrivial quantum theory in the limit of infinite cutoff (it was shown afterwards that a different choice of the parameters that involve a wave function renormalization 
 is also possible, resulting in a completely different quantum theory \cite{autonomous}). Both possibilities have their corresponding counterparts in curved spacetimes \cite{mazzi-paz}.  
The quantum theories so derived have different drawbacks, which originate from the fact that in the perturbative regime (where the renormalized coupling 
constants are very small) those theories do not approach the ones defined by the usual perturbation theory in terms of Feynman diagrams,
which are based on a loop-by-loop renormalization procedure. 
Moreover, for finite values of the cutoff the theory is unstable, since the bare coupling constant is negative and goes to zero as the cutoff tends to infinite
\cite{Stevenson,Amelino}. In addition, there is no clear way to systematically improve the approximation based on such variational principle. 

The use of the 2PI EA solves some of these drawbacks by changing the renormalization procedure. Indeed, in this formalism, there are  multiple definitions of
a given $n$-point function, which are equivalent in the exact theory but not necessarily  if some approximation is considered. When the theory is truncated,
it is mandatory to impose some consistency relations between the different counterterms, and this is possible only if one allows for more bare constants than usual,
which are to be fixed in terms of the usual number of renormalized  parameters (the ones that are in principle measurable). In other words, for a given 
renormalized parameter one includes different counterterms. There has been much progress in understanding the renormalization properties of general 
truncation of  the 2PI effective action, and a systematic method for a consistent renormalization has been developed \cite{Bergesetal}.

In this paper we will analyze interacting quantum fields in general spacetimes using the 2PI EA with the consistent renormalization procedure of 
Ref. \cite{Bergesetal}. We will show that the renormalization procedure can be generalized to curved spacetimes, and we will find the explicit renormalized 
form for the mean value equation and for the propagator of the fluctuations. We will then analyze the particular case of de Sitter spacetime. The study of the mean value equation along with the corresponding effective potential will allow us 
to show that, under this nonperturbative 
approximation and renormalization, depending on the values of the parameters, spontaneous symmetry breaking may take place in de Sitter spacetime. This is
to be contrasted with the results based 
on the approach of Ref. \cite{Stevenson}, in which one can prove that spontaneous symmetry breaking is  not compatible with de Sitter symmetries \cite{mazzi-paz} (a similar incompatibility holds to leading order in the $1/N$ approximation to $O(N)$ models, as mentioned in
\cite{mazzi-paz}; see also \cite{Serreau0}). 
Our results show discrepancies with those of Ref.\cite{Arai}, where the effective potential in de Sitter spacetime has been also recently 
considered using the consistent renormalization procedure. As we will see, the discrepancies come from the milder version of the consistency
conditions used in that work.

Before describing the details of the calculations, it is worth to stress here that  the aim of this paper is not  to arrive at a conclusive answer 
to these points. Indeed,  previous works based in the Hartree
approximation in flat spacetime suggest that this is not  possible. For example, it is well known that the Hartree approximation predicts a first 
order phase transition when
applied to $O(N)$ models, while other nonperturbative methods give a second-order phase transition \cite{firstorder}, in agreement with
expectations based on general arguments \cite{Tetradis}. These drawbacks of the Hartree approximation can be cured by including
two loops corrections to the 2PI EA \cite{Marko}. A similar situation could arise in curved spacetimes, in particular when discussing 
symmetry breaking in de Sitter spacetime. 
Therefore, one of our purposes here is to push forward a rigorous and critical  analysis of the situation,
by starting with an understanding of the simplest nonperturbative approximation, along with its limitations, 
but having in mind the necessity  of an improvement.  This subject is clearly worth studying in detail, to the end of analyzing its 
eventual relevance not only for correlation functions, but also for the evolution of the Universe itself through its contribution to the  
energy momentum tensor in the Semiclassical Einstein Equations (SEE), which is the subject of a forthcoming  paper II \cite{Nos2}.

The paper is organized as follows. In Sec. \ref{2PISect} we introduce the 2PI EA in curved spacetimes, and make contact with previous formulations based on the use
of the equations for the mean value of the field and its  two-point function. In Sec. \ref{RenGapSect} we discuss in detail the renormalization of the field and gap 
equations, using the consistent renormalization procedure of \cite{Bergesetal}.  We show that the renormalized equations are independent of the scale 
$\tilde\mu$ introduced by the dimensional regularization, once we express the equations in terms of the renormalized mass and coupling constants defined in Minkowski spacetime.
In Sec. \ref{DSSect} we analyze the field and gap equations in a de Sitter spacetime, exploring the possibility of having spontaneous symmetry breaking. In Sec. \ref{RP} we
generalize all the previous results to the case in which the renormalization point is taken for a given fixed de Sitter metric.  This change of the renormalization point
 allows us  to explore further the parameter space where spontaneous symmetry breaking is possible.
In Sec. \ref{ConcluSect} we present a summary and a discussion of the results obtained in the paper. The Appendices contain some details of the calculations.
Everywhere we set $c=\hbar=1$ and adopt the mostly plus sign convention.

\section{The 2PI effective action}\label{2PISect}

The definition of the 2PI EA along with the corresponding functional integral can be found in both  papers and textbooks (see for instance
\cite{calzetta,CJT,Ramsey}). In this section,
we briefly  summarize the main relevant aspects of the formalism for the case of  a self-interacting scalar field $\phi$ in a general curved spacetime. 
 
The 2PI generating functional can be written as \cite{Bergesetal} 
\begin{equation}
 \Gamma_{2PI}[\phi_0,G,g^{\mu \nu}] = S_0[\phi_0,g^{\mu \nu}] + \frac{i}{2} \textup{Tr} \ln(G^{-1}) + \frac{i}{2} \textup{Tr}( G_0^{-1} G) + \Gamma_{int}[\phi_0,G,g^{\mu \nu}],
\label{gamma-2PI-alt}
\end{equation}
where $S_0$ is quadratic part of the classical action $S$ without any counterterms and
\begin{equation}
i G_0^{ab}(x,x') = \frac{1}{\sqrt{-g}} \frac{\delta^2 S_0[\phi_0,g^{\mu\nu}]}{\delta \phi_a(x) \delta \phi_b(x')} \frac{1}{\sqrt{-g'}},
\end{equation} with $a, b$  the time branch indices (with index set $\{+,-\}$ in the usual notation) corresponding to the ordering on the contour in  the  ``closed-time-path''(CTP) or  Schwinger-Keldysh  \cite{calzetta} formalism. This formalism is in principle needed in order to obtain real and causal evolution equations, although
in the approximation we will use in this paper the details of  the ordering along the contour will not be  needed.

The interaction part of the 2PI EA is given by
\begin{equation}
 \Gamma_{int}[\phi_0,G,g^{\mu \nu}] = S_{int}[\phi_0,g^{\mu \nu}] + \frac{1}{2} \Tr \left[ \frac{\delta^2 S_{int}}{\delta \phi_0 \delta \phi_0} G \right] + \Gamma_2[\phi_0,G,g^{\mu \nu}],
\end{equation} where $S_{int} = S - S_0$, the functional $\Gamma_2$ is $(-i)$ times the sum of all two-particle-irreducible vacuum-to-vacuum diagrams with lines given by $G$ and vertices obtained from the shifted action $S^{F}_{int}$, which comes from expanding $S_{int}[\phi_0 + \varphi]$ and collecting all terms higher than quadratic in the fluctuating field $\varphi$.

The equations of motion are obtained by extremizing the effective action
\begin{subequations}
\begin{align}
 \frac{\delta \Gamma_{2PI}}{\delta \phi_0}\Big{|}_{\phi_{+}=\phi_{-}=\phi; g^{\mu\nu}_{+}=g^{\mu\nu}_{-}=g^{\mu\nu}} &= 0,\\
 \frac{\delta \Gamma_{2PI}}{\delta G}\Big{|}_{\phi_{+}=\phi_{-}=\phi;g^{\mu\nu}_{+}=g^{\mu\nu}_{-}=g^{\mu\nu}}  &= 0.
\end{align}
\end{subequations}
The usual 1PI EA is then obtained by replacing the solution of the gap equation $\bar{G}[\phi_0]$ back in the 2PI EA
\begin{equation}
 \Gamma_{1PI}[\phi_0,g^{\mu \nu}] = \Gamma_{2PI}[\phi_0,\bar{G}[\phi_0],g^{\mu \nu}].
 \label{1PI-2PI}
\end{equation}

As mentioned in the introduction, in the 2PI formalism there are various possible $n$-point functions that can be obtained from the several possible ways
of functionally differentiating $\Gamma_{2PI}[\phi_0,G,g^{\mu\nu}]$ with respect to $\phi_a$ and $G_{ab}$. However in the exact theory it can be
shown that the different $n$-point functions are not independent, and must satisfy certain consistency conditions. These are thoroughly derived 
in  \cite{Bergesetal}, of which we provide a brief sketch in  Appendix A. To the purposes of this paper, it is enough to consider a simplified 
version of these conditions which are only valid for a theory with $Z_2$ symmetry, and taking advantage of further simplifications when evaluating 
at $\phi_0=0$. The two important conditions in our case are
\begin{equation}
  \frac{\delta^2 \Gamma_{int}}{\delta \phi_1 \delta \phi_2} \Bigg|_{\phi=0} = 2 \frac{\delta \Gamma_{int}}{\delta G_{12}} \Bigg|_{\phi=0},
\label{2-pt-relation}
\end{equation}
and 
\begin{eqnarray}
\frac{\delta^4 \Gamma_{1PI}[\phi_0]}{\delta \phi_1 \delta \phi_2 \delta \phi_3 \delta \phi_4}\Bigg|_{\phi_0=0} = 2 \left[ \frac{\delta^2 \Gamma_{int}}{\delta G_{12} \delta G_{34}} \Bigg|_{\bar{G},\phi_0=0} + perms(2,3,4) \right] - \frac{1}{2} \frac{\delta^4 \Gamma_{int}}{\delta \phi_1 \delta \phi_2 \delta \phi_3 \delta \phi_4} \Bigg|_{\bar{G},\phi_0=0}, \,\,\,\,\,\,
 \label{4-pt-relation}
\end{eqnarray}
where expression \eqref{1PI-2PI} is used to establish a connection between the 1PI effective action on the left hand side and $\Gamma_{int}$ on the right hand side.

It is worth to recall that, due to an artifact of the approximation, these conditions  can be violated  when applied to a particular truncation of  $\Gamma_{2PI}$.
However, if the approximation is good any departure from these conditions should be small. Nevertheless,  since we are dealing with formally divergent 
quantities,  the latter statement is non-trivial, and therefore a consistent renormalization procedure that takes these relations into account is required. 
In fact, an essential step in the consistent renormalization procedure of Ref. \cite{Bergesetal} is to impose these conditions to the particular truncation of 
$\Gamma_{2PI}$ at a given renormalization point (for which we choose $\phi_0=0$), which allows to fix the different counterterms.

\subsection{$\lambda \phi^4$ theory in the Hartree approximation}

\begin{figure}

\setlength{\unitlength}{1.5cm}
\begin{picture}(3,0)
\put(0,0){\circle{5}}
\put(-0.48,0){\line(1,0){0.95}}
\put(2,0){\circle{5}}
\put(2.94,0){\circle{5}}
\end{picture}
\vspace{0.5cm}

\caption{ 2PI  diagrams at 2-loop order: the ``sunset'' (on the left) and the ``double-bubble'' (to the right).}\label{diagrams}
\end{figure}
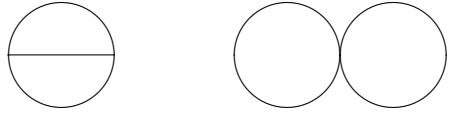

For a non-minimally coupled scalar field with quartic self-coupling in a curved background with metric $g_{\mu \nu}$ the classical action reads
\begin{equation}
 S[\phi,g^{\mu\nu}] = -\int d^4 x \, \sqrt{-g} \left[ \frac{1}{2} \phi \left( -\square + m^2_B + \xi_B R \right) \phi + \frac{1}{4!} \lambda_B \phi^4 \right],
 \label{classical-S}
\end{equation}
where $\square = \frac{1}{\sqrt{-g}} \partial_\mu \left( \sqrt{-g} g^{\mu \nu} \partial_\nu \right)$, $g \equiv \det (g_{\mu\nu})$. Then the shifted action that defines the interaction vertices needed to construct the 2-particle irreducible vacuum diagrams is
\begin{equation}
S^{F}_{int}[\varphi,\phi_0,g^{\mu\nu}]  = - \frac{\lambda_B}{6} \int d^4 \,x \, \sqrt{-g} \left[ \frac{1}{4} \varphi^4 + \phi_0 \varphi^3 \right]. 
\label{shifted-action}
\end{equation}
Although we are dealing with a scheme in which there is an infinite resummation of diagrams, we cannot compute the effective action completely because we still have to perform an infinite sum of 2PI vacuum diagrams in the $\Gamma_2$ term. Therefore we must resort to some kind of approximation. At the lowest order one can drop the $\Gamma_2$ term altogether, which corresponds to the 1-loop approximation in which case there is no difference with the 1PI effective action. This is because the 2PI vacuum diagrams start at 2-loops, and it is from this order onwards that the 2PI effective action gives a nontrivial result compared to the 1PI effective action. With the shifted action $S_{int}$ given by Eq. \eqref{shifted-action}, there are two diagrams that contribute at the 2-loop order. These are the double-bubble and the sunset shown in Fig. \ref{diagrams}. 

The Hartree approximation corresponds to taking into account only the local contribution (double-bubble diagram). In this case the 2PI effective action is
\begin{eqnarray}
 \Gamma_{2PI}[\phi_0,G,g_{\mu\nu}] &=& -\int d^4 x \, \sqrt{-g} \left[ \frac{1}{2} \phi_0 \left( -\square + m_{B2}^2 + \xi_{B2} R  \right) \phi_0 + \frac{1}{4!} \lambda_{B4} \phi_0^4 \right] + \frac{i}{2} \Tr \ln ( G^{-1} ) \nonumber \\ 
&&- \frac{1}{2} \int d^4 x \, \sqrt{-g} \left[ -\square + m_{B0}^2 + \xi_{B0} R + \frac{1}{2} \lambda_{B2} \phi_0^2 \right] G(x,x) \label{2PI-lambdaphi4} \\
\nonumber
&& - \frac{\lambda_{B0}}{8} \int d^4 x \, \sqrt{-g} \, G^2(x,x).  
\end{eqnarray}
Note that in each contribution to the effective action we allow the counterterms to be different, denoted as different subscripts in the bare parameters
that refer to the power of $\phi_0$ in the corresponding term of the action. This allows for the possibility of adjusting how each diagram contributes to
the cancellation of divergences, which turns out to be a crucial point for the renormalization procedure to respect the 2PI consistency relations of the 
exact theory. The relationship between the different counterterms will be fixed by imposing the consistency relations on the 2 and 4-point functions 
in the next section.  It is important to note that, within the Hartree approximation, the CTP formalism gives the 
same equations of motion than the usual {\it in-out} formalism \cite{Ramsey}. Therefore, in Eq. (\ref{2PI-lambdaphi4}) and in what follows we drop the time branch indices.  

The equations of motion for the mean field and the exact propagator (gap equation) are obtained by taking the variation of the $\Gamma_{2PI}[\phi_0,G]$ with respect to $\phi_0$ and $G$ respectively. In the Hartree approximation that we are considering  these read
\begin{eqnarray}
 \left( -\square + m_{B2}^2 + \xi_{B2} R + \frac{\lambda_{B4}}{6}  \phi_0^2 + \frac{\lambda_{B2}}{2}  G(x,x) \right) \phi_0(x) &=& 0, \\
 \left( -\square + m^2_{B0} + \xi_{B0} R + \frac{\lambda_{B2}}{2}  \phi_0^2 + \frac{\lambda_{B0}}{2}  G(x,x) \right) G(x,x') &=& -i \frac{\delta(x-x')}{\sqrt{-g'}}.
\end{eqnarray}
These equations are similar to those obtained when considering a Gaussian approximation at the level of the mean field equations \cite{mazzi-paz}. In that case
the starting point is the classical equation of motion for $\phi$, which is then separated into a mean field $\phi_0 = \langle \phi \rangle$ and a
fluctuation $\varphi = \phi - \phi_0$. Then taking the expectation value of the classical field equation one gets a pair of coupled equations for 
$\phi_0$ and $\varphi$, or equivalently for $\phi_0$ and the propagator of the fluctuations $G(x,x')$. Finally upon the assumption of Gaussian states,
which implies
\begin{subequations}
\begin{align}
 \langle \varphi^3 \rangle &= 0, \\
 \langle \varphi^4 \rangle &= 3\langle \varphi^2 \rangle^2,
\end{align}
\end{subequations}
the resulting equations are similar to those shown above. The difference is, however, that there is neither a diagrammatic interpretation 
for each of the contributions that allows for different counterterms, nor a set of consistency relations to fix them.
Therefore in this approach it would seem unnatural to use different counterterms.


\section{Renormalization of the field and gap equations}\label{RenGapSect}

In the analysis of the renormalization of the Hartree approximation, we will use the following parametrization of the bare couplings:
\begin{subequations}
\begin{align}
 m^2_{Bi} &= m^2 + \delta m_i, \\
 \xi_{Bi} &= \xi + \delta \xi_i, \\
 \lambda_{Bi} &= \lambda + \delta \lambda_i, 
\end{align}
\label{type-of-counterterms}
\end{subequations}
which corresponds to the MS scheme (i.e., the counterterms $\delta m_i$, $\delta \xi_i$ and $\delta \lambda_j$ ($i=0,2$,$j=0,2,4$) contain only divergences and no finite part).
As mentioned in the previous section, in order to fix the different counterterms within this level of approximation we impose
the consistency relations \eqref{2-pt-relation} and \eqref{4-pt-relation} on the different 2PI 2 and 4-point kernels. These kernels are 
computed from functional derivatives of $\Gamma_{int} \left[ \phi_0, G \right]$ evaluated at $\phi_0=0$, which by comparison of 
\eqref{gamma-2PI-alt}  and \eqref{2PI-lambdaphi4} reads
\begin{eqnarray}
 \Gamma_{int}[\phi_0,G,g_{\mu\nu}] &=& -\int d^4 x \, \sqrt{-g} \left[ \frac{1}{2} \left( \delta m_2 + \delta \xi_2 R \right) \phi_0^2 + \frac{1}{4!} \left( \lambda + \delta \lambda_4 \right) \phi_0^4 \right] \notag \\ 
&&- \frac{1}{2} \int d^4 x \, \sqrt{-g} \left[ \delta m_0 + \delta \xi_0 R + \frac{1}{2} \left( \lambda + \delta \lambda_2 \right) \phi_0^2 \right] G(x,x) \label{gammaint-lambdaphi4} \\
&& - \frac{1}{8} \int d^4 x \, \sqrt{-g} \left( \lambda + \delta \lambda_0 \right) G^2(x,x),  \notag
\end{eqnarray}
where we have used that the inverse free propagator $G_0^{-1}$ in \eqref{gamma-2PI-alt} is defined as
\begin{equation}
 G_0^{-1} = i \left( -\square + m^2 + \xi R \right).
\end{equation}
After some straightforward functional differentiations,  the 2-point kernels at $\phi_0 = 0$ are given by
\begin{eqnarray}
 \frac{\delta^2 \Gamma_{int}}{\delta \phi_0(x) \delta \phi_0(x')} \Bigg|_{\phi_0=0} &=& - \sqrt{-g} \left[ \delta m_2 + \delta \xi_2 R + \frac{1}{2} (\lambda + \delta \lambda_2 ) G(x,x) \right] \delta(x-x'), \\
2 \frac{\delta \Gamma_{int}}{\delta G(x,x')} \Bigg|_{\phi_0=0} &=& - \sqrt{-g}\left[ \delta m_0 + \delta \xi_0 R + \frac{1}{2}(\lambda + \delta \lambda_0 ) G(x,x) \right] \delta(x-x').
\end{eqnarray}
Hence the condition \eqref{2-pt-relation} implies that
\begin{subequations}
\begin{align}
 \delta m_0 &= \delta m_2 \equiv \delta m, \label{masses} \\
 \delta \xi_0 &= \delta \xi_2 \equiv \delta \xi, \label{xis} \\
 \delta \lambda_0 &= \delta \lambda_2. \label{lambdas0-2}
\end{align}
\end{subequations}
Since both mass counterterms are equal, as also both curvature coupling counterterms are, we drop the subindexes in those cases from now on.

We turn to the second consistency relation Eq. \eqref{4-pt-relation}. The equation relates the different 4-point kernels at $\phi_0 = 0$, two of which can be calculated by carrying on differentiating $\Gamma_{int} \left[ \phi_0, G \right]$:
\begin{eqnarray}
 \frac{\delta^4 \Gamma_{int}}{\delta \phi_1 \delta \phi_2 \delta \phi_3 \delta \phi_4} \Bigg|_{\phi_0=0} &=& - \sqrt{-g} \, (\lambda + \delta \lambda_4 ) \delta_{12} \delta_{13} \delta_{14}, \\
 \frac{\delta^2 \Gamma_{int}}{\delta G_{12} \delta G_{34}} \Bigg|_{\phi_0=0} &=& - \frac{\sqrt{-g}}{4}(\lambda + \delta \lambda_0 ) \delta_{12} \delta_{34} \delta_{23},
\end{eqnarray}
where we used $\phi_i \equiv \phi_0(x_i)$ as a notational shorthand. The other quantity that must be specified is the fourth functional derivative of the 1PI effective action. Its value at $\phi_0=0$ is then easily interpreted as the renormalized self-interaction coupling $\lambda_R$, since the effective potential is proportional to the effective action at a constant value of $\phi_0$. Therefore
\begin{equation}
 \frac{\delta^4 \Gamma_{1PI}[\phi_0]}{\delta \phi_1 \delta \phi_2 \delta \phi_3 \delta \phi_4} \Bigg|_{\phi_0=0} = - \sqrt{-g} \, \lambda_R \delta_{12} \delta_{13} \delta_{14}, 
\end{equation}
is equivalent to 
\begin{equation}
\lambda_R = \frac{d^4 V_{eff}}{d \phi_0^4} \Bigg|_{\phi_0=0}. \label{lambda-def}
\end{equation}
It is important to note that, since the 1PI effective action depends on the background geometry, in general it is  not possible  to impose these consistency conditions exactly for any arbitrary background metric.
We expect   an analogous situation to  happen  in Minkowski space if for instance the background field is time dependent. However, in that case, in order to define the renormalized parameters, one 
can still  impose the consistency conditions for a particular constant value of the field. This is enough to fix the counterterms. Then, once made finite, 
 any deviation from the consistency conditions is expected to be small, or more precisely to be of the same order as the contributions neglected 
 in the approximation considered for $\Gamma_{1PI}$. The corresponding choice in our case, is to consider a spacetime with constant curvature. We will come back to this point in Sec. \ref{RP}.  

Feeding the above ingredients to the second consistency relation \eqref{4-pt-relation} fixes the remaining counterterm. Here it is important to take into account the symmetry properties of the kernels to evaluate the different permutations. The resulting relation is
\begin{equation}
 \delta \lambda_4 - 3 \delta \lambda_2 = 2( \lambda - \lambda_R).
 \label{lambdas2-4}
\end{equation}
At this point one can see more clearly the consequences of this ``arbitrary'' truncation of the 2PI effective action, i.e. the Hartree approximation, if we insist on enforcing the properties that are valid for the exact theory. 

Turning back to the renormalization of the mean field and gap equations, instead of considering the equation for the Feynman propagator $G(x,x')$, we will consider the equation for the Hadamard propagator $G_1(x,x') = \langle \{ \varphi(x) , \varphi(x') \} \rangle$, which contains the same information but is more convenient for our purposes 
\begin{equation}
 \left( -\square + m^2_{B0} + \xi_{B0} R + \frac{\lambda_{B2}}{2}  \phi_0^2 + \frac{\lambda_{B0}}{4} [G_1] \right) G_1(x,x') = 0,
\end{equation}
where $[G_1] = G_1(x,x) = 2 G(x,x) = 2 \langle \varphi^2 \rangle$. Renormalizability of the mean field and gap equations means that they can be rendered finite by a suitable choice of counterterms. If this is the case, the resulting equations can be expressed in terms of a finite physical mass plus a coupling to the curvature term, namely
\begin{eqnarray}
 \left( -\square + m_{ph}^2 + \xi_R R - \frac{1}{3} \lambda_R \phi_0^2 \right) \phi_0(x) &=& 0, \label{field-eq} \\ 
 \left( -\square + m_{ph}^2 + \xi_R R \right) G_1(x,x') &=& 0.\label{gap-eq}
\end{eqnarray}
The physical mass $m_{ph}^2$ is a spacetime dependent scalar function determined by
\begin{equation}
 m_{ph}^2 + \xi_R R = m^2 + \delta m + (\xi + \delta \xi) R + \frac{1}{2} (\lambda + \delta \lambda_2) \phi_0^2 + \frac{1}{4} (\lambda + \delta \lambda_2) [G_1],
 \label{phys-mass}
\end{equation}
which is a self-consistent equation, as $m_{ph}^2$ also enters the right hand side through $[G_1]$. In this expression we have already used the relationships among the counterterms as dictated by the consistency conditions. The divergences come from $[G_1]$,  and must be cancelled by a suitable choice of the counterterms $\delta m$, $\delta \xi$ and $\delta \lambda_2$. 

To the end of exposing and isolating the divergences, we use a Schwinger-DeWitt type expansion for $[G_1]$ for a free field with variable mass $m_{ph}$ and coupling to the curvature $\xi_R$ on a general $n$-dimensional background with metric $g_{\mu\nu}$:
\begin{eqnarray}
 [G_1] &=& \frac{1}{8 \pi^2} \left( \frac{m_{ph}^2}{\mu^2} \right)^{\epsilon/2} \sum_{j \geq 0} \, [\Omega_j] (m_{ph}^2)^{1-j} \, \Gamma\left( j-1 - \frac{\epsilon}{2} \right) \notag\\
  &\equiv& \frac{1}{4\pi^2 \epsilon} \left[ m_{ph}^2 + \left(\xi_R - \frac{1}{6} \right) R \right] + 2\, T_F (m_{ph}^2,\xi_R,R,\tilde{\mu}). \label{adiab-G1} 
\end{eqnarray}
Here $\epsilon = n - 4$, $\Gamma(x)$ is the Gamma function and the coefficients $[\Omega_j]$ are scalars of adiabatic order $2j$ built from the metric and its derivatives and satisfy certain recurrence relations. We use the results for these coefficients presented in \cite{mazzi-paz2}. In the second line we have split the sum by isolating the zeroth and second adiabatic orders, which are the only divergent terms. Then we expanded for $\epsilon \to 0$ and redefined $\mu \to \tilde{\mu}$ to absorb some constant terms. The finite part is
\begin{eqnarray}
\nonumber T_F (m_{ph}^2,\xi_R,R,\tilde{\mu})& =& \frac{1}{16 \pi^2}  \Bigg{\{} \left[ m_{ph}^2 + \left(\xi_R - \frac{1}{6} \right) R \right] \ln \left( \frac{m_{ph}^2}{\tilde{\mu}^2} \right) + \left(\xi_R - \frac{1}{6} \right) R \\ 
&- & 2 F(m_{ph}^2,\{R\}) \Bigg{\}}
\end{eqnarray}
where the function $F(m_{ph}^2,\{R\})$ contains the adiabatic orders higher than two and it is independent of $\epsilon$ and $\mu$. The dependence on $m_{ph}^2$ involves also its derivatives, while that denoted by $\{R\}$ is to be understood as a dependence on curvature invariants constructed from contractions of the Riemann tensor and its derivatives. This function satisfies the following properties
\begin{subequations}
\begin{align}
&F(m_{ph}^2,\{R\})\bigg|_{R_{\mu\nu\rho\sigma} = 0} = 0, \label{F-prop1} \\
&\frac{dF(m_{ph}^2,\{R\})}{d m_{ph}^2} \Bigg|_{R_{\mu\nu\rho\sigma}=0} = 0,\label{F-prop2} \\
&\frac{dF(m_{ph}^2,\{R\})}{d R} \Bigg|_{R_{\mu\nu\rho\sigma}=0} = 0. \label{F-prop3}
\end{align}
\label{F-props}
\end{subequations}
Then inserting Eq. \eqref{adiab-G1} back in Eq. \eqref{phys-mass}, we have
\begin{eqnarray}\nonumber
 m_{ph}^2 + \xi_R R& =& m^2 + \delta m + (\xi + \delta \xi) R + \frac{1}{2} (\lambda + \delta \lambda_2) \phi_0^2 + \frac{1}{16\pi^2 \epsilon} (\lambda + \delta \lambda_2) \left[ m_{ph}^2 + \left(\xi_R - \frac{1}{6} \right) R \right]  \\
 &+& \frac{1}{2} (\lambda + \delta \lambda_2) T_F.
\end{eqnarray}
To separate unambiguously the divergent part from the finite part it is necessary to choose a subtraction scheme. In the MS scheme, the divergent part must vanish, and then the following equations are satisfied independently
\begin{eqnarray}
 &&m_{ph}^2 + \xi_R R = m^2 + \xi R + \frac{1}{2} \lambda \phi_0^2 + \frac{1}{2} \lambda T_F, \\
 &&0 = \Bigg\{ \delta m + \delta \xi R + \frac{1}{2} \delta \lambda_2 \phi_0^2 + \frac{1}{16\pi^2 \epsilon} (\lambda + \delta \lambda_2) \left[ m_{ph}^2 + \left(\xi_R - \frac{1}{6} \right) R \right] + \frac{1}{2} \delta \lambda_2 T_F \Bigg\}.
\end{eqnarray}
These conditions will determine the MS counterterms $\delta m$, $\delta \xi$ and $\delta \lambda_2$. Using the first equation to express the $m_{ph}^2 + \xi_R R$ and replace it in the second, we end up with an expression that depends on the counterterms and the finite constants
\begin{eqnarray}
0& =& \left[ \delta m + \frac{m^2}{16\pi^2 \epsilon} (\lambda + \delta \lambda_2) \right] + \left[ \delta \xi + \frac{1}{16\pi^2 \epsilon} \left( \xi - \frac{1}{6} \right) (\lambda + \delta \lambda_2) \right] R\nonumber \\
&+& \frac{1}{2} \left[ \delta \lambda_2 + \frac{\lambda}{16\pi^2 \epsilon} (\lambda + \delta \lambda_2) \right] \left( \phi_0^2 + T_F \right),
\end{eqnarray}
where each pair of square brackets must vanish independently. The resulting counterterms are 
\begin{subequations}
\begin{align}
\delta m &= -\frac{\lambda}{16\pi^2 \epsilon} \left( \frac{m^2}{1 + \frac{\lambda}{16\pi^2 \epsilon}} \right), \\
\delta \xi &= -\frac{\lambda}{16\pi^2 \epsilon} \left( \frac{\left( \xi - \frac{1}{6} \right)}{1 + \frac{\lambda}{16\pi^2 \epsilon}} \right), \\
\delta \lambda_2 &= -\frac{\lambda}{16\pi^2 \epsilon} \left( \frac{\lambda}{1 + \frac{\lambda}{16\pi^2 \epsilon}} \right), 
\end{align}
\end{subequations}
and the bare parameters become
\begin{subequations}
\begin{align}
m_B^2 &= \frac{m^2}{1+\frac{\lambda}{16\pi^2\epsilon}} = m^2\sum_{n=0}^{+\infty} \left(-\frac{\lambda}{16\pi^2\epsilon}\right)^n, \label{mB}\\
\xi_B-\frac{1}{6} &= \frac{\left(\xi-\frac{1}{6}\right)}{1+\frac{\lambda}{16\pi^2\epsilon}} = \left(\xi-\frac{1}{6}\right)\sum_{n=0}^{+\infty} \left(-\frac{\lambda}{16\pi^2\epsilon}\right)^n, \label{xiB} \\
\lambda_{B2} &= \frac{\lambda}{1+\frac{\lambda}{16\pi^2\epsilon}} = \lambda\sum_{n=0}^{+\infty} \left(-\frac{\lambda}{16\pi^2\epsilon}\right)^n. \label{lambdaB}
\end{align}
\label{bare-couplings}
\end{subequations}


Once made finite, the equation for the physical mass is
\begin{eqnarray}
 m_{ph}^2 + \xi_R R &=& m^2 + \xi R + \frac{1}{2} \lambda \phi_0^2 + \frac{\lambda}{32\pi^2} \left\{ \left[ m_{ph}^2 + \left(\xi_R - \frac{1}{6} \right) R \right] \ln \left( \frac{m_{ph}^2}{\tilde{\mu}^2} \right) \right.\nonumber\\
 &+&\left. \left(\xi_R - \frac{1}{6} \right) R - 2 F(m_{ph}^2,\{R\}) \right\}. 
 \label{phys-mass-fin}
\end{eqnarray}
This is a self-consistent equation for $m_{ph}^2(\phi_0,R)$, whose result shall then be inserted into the field equation \eqref{field-eq} and solved for $\phi_0$. Expressed in this way, the result depends on a mixture of the finite MS-parameters $m^2$, $\xi$ and $\lambda$ and the renormalized parameters $\xi_R$ and $\lambda_R$ (this last one coming from the consistency condition through the field equation), as well as in the regularization scale $\tilde{\mu}$. The renormalized parameters are those that characterize the effective potential $V_{eff}$ and will not be equal to the MS-parameters in general. Both sets of parameters will be related by $\tilde{\mu}$. It is convenient then to express the $m_{ph}^2$-equation in terms of only one set of parameters, for which we use the effective potential as identified through its derivative in the field equation \eqref{field-eq}
\begin{equation}
 \frac{d V_{eff}}{d \phi_0} = \left(m_{ph}^2 + \xi_R R - \frac{1}{3} \lambda_R \phi_0^2 \right) \phi_0,
 \label{Veff}
\end{equation}
in order to find relations between both sets. As it was discussed in the previous section, the consistency conditions take a particularly simple form only at $\phi_0 = 0$, so we will use that choice of renormalization point. It is also necessary to fix the background geometry at a constant curvature. In this section we will choose Minkowski spacetime as the renormalization point, for which $R=0$. With these to conditions we define the renormalized parameters as
\begin{subequations}
\begin{align}
m_R^2 &\equiv \frac{d^2 V_{eff}}{d \phi_0^2} \Bigg|_0 = m_{ph}^2|_0, \label{ren-mass-def} \\
\xi_R &\equiv \frac{d^3 V_{eff}}{d R \, d \phi_0^2} \Bigg|_0 = \frac{d m_{ph}^2}{d R} \Bigg|_0 + \xi_R, \label{ren-xi-def} \\
\lambda_R &\equiv \frac{d^4 V_{eff}}{d \phi_0^4} \Bigg|_0 = 3 \frac{d^2 m_{ph}^2}{d \phi_0^2} \Bigg|_0 - 2 \lambda_R, \label{ren-lambda-def}
\end{align}
\label{ren-quantities}
\end{subequations}
where the $0$ subindex means both zero curvature and $\phi_0 = 0$. The definition \eqref{ren-lambda-def} is the same as \eqref{lambda-def}. From these definitions and Eqs. \eqref{Veff} and \eqref{phys-mass-fin}, it is straightforward to arrive at expressions that relate the renormalized parameters $m_R^2$, $\xi_R$ and $\lambda_R$ to the MS finite parameters $m^2$, $\xi$ and $\lambda$ and $\tilde{\mu}$ (see Appendix B). The resulting relations are
\begin{subequations}
\begin{align}
m_R^2 &= \frac{m^2}{\left[ 1 - \frac{\lambda}{32 \pi^2} \ln \left( \frac{m_R^2}{\tilde{\mu}^2} \right) \right]}, \label{mR-mu} \\
\left( \xi_R - \frac{1}{6} \right) &= \frac{\left( \xi - \frac{1}{6} \right)  }{ \left[1 - \frac{\lambda}{32\pi^2} - \frac{\lambda}{32\pi^2} \ln \left( \frac{m_R^2}{\tilde{\mu}^2} \right) \right] }, \label{xiR-mu} \\
 \lambda_R &= \frac{\lambda}{\left[ 1 - \frac{\lambda}{32\pi^2} - \frac{\lambda}{32\pi^2} \ln \left( \frac{m_{R}^2}{\tilde{\mu}^2} \right) \right]}. \label{lambdaR-mu}
\end{align}
\label{ren-mu}
\end{subequations} 
Putting these together we can find some useful $\tilde{\mu}$-independent combinations of the MS parameters:
\begin{equation}
 \frac{\left( \xi_B - \frac{1}{6} \right)}{\lambda_B} = \frac{\left( \xi - \frac{1}{6} \right)}{\lambda} = \frac{\left( \xi_R - \frac{1}{6} \right)}{\lambda_R},
 \label{xi-lambda}
\end{equation}
and
\begin{equation}
 \frac{m_B^2}{\lambda_{B2}} = \frac{m^2}{\lambda} = m_R^2 \left( \frac{1}{32\pi^2} + \frac{1}{\lambda_R} \right) \equiv \frac{m_R^2}{\lambda_R^{*}},
 \label{m-lambda}
\end{equation}
where we have introduced $\lambda_R^*$ as a notational shorthand, with the property $\lambda_R^* \to \lambda_R$ for $\lambda_R \ll 1$.

Note that, although in principle $m^2$ could take negative values, the parameter $m_R^2$ is positive by construction. 
This is so because  $m_R^2$ is a solution of the gap equation in Minkowski space at $\phi_0=0$, which is based on the existence of an stable propagator for the fluctuations (see Eq. (\ref{adiab-G1})). Therefore,  these equations tell us that in this case $m^2$ must also be positive, provided that $\lambda_R^*$ and $\lambda$ are positive. This is a consequence of having defined the renormalized parameters in Minkowski spacetime, and as we will see in Sec. V, this restriction can be relaxed by taking the renormalization point in de Sitter spacetime.

 In any case, it is worth to remark that the consistency conditions impose nontrivial constraints on the finite parameters of the theory. These restrictions are not considered (or at least not apparently) in the approach of Ref. \cite{Arai} where, working with minimal subtraction, it is assumed from the beginning that  $\delta\lambda_4=3\delta\lambda_2$.  On  one hand, as we are working with an approximation to  the EA,  one could argue that it is not necessary to impose  the consistency conditions exactly at the renormalization point, but only to use the conditions to fix the proportionality constant between the counterterms, because anyway they are  not expected to be exact beyond that point. This is in principle correct. However, unless a set of renormalization conditions is specified, the interpretation of the finite parameters  is unclear and the  equations are   $\tilde\mu$-dependent, as happen in Ref. \cite{Arai}. Moreover, if we assume that   $\delta\lambda_4=3\delta\lambda_2$,  and then we define the renormalization conditions as derived from the effective potential (as we are doing), it turns out that the gap equation cannot be entirely written in terms of only the renormalized parameters, and hence it is $\tilde\mu$-dependent. This can be easily seen by noticing that it is due to the combination on the rhs of Eq.   (\ref{lambdas2-4}) that it is $\lambda_R$ and not $\lambda$ which appears  in Eq. (\ref{field-eq}).  On the other hand, if the consistency conditions are imposed, the  relation $\delta\lambda_4=3\delta\lambda_2$  implies  that  $\lambda=\lambda_R$, which yields a particular choice of the parameter $\tilde\mu$. Our analysis shows that, taking Minkowski spacetime as the renormalization point, the choice  $m^2<0$  is not compatible with the consistency conditions. Similar restrictions will appear when considering a more general definition for the renormalized parameters (see Section V).

The  above relations between the parameters can be used to rewrite the equation for $m_{ph}^2$ in terms of the renormalized parameters only. After a bit of algebra we arrive at
\begin{eqnarray}
  m_{ph}^2 &=& m_R^2 + \frac{\lambda_R^*}{2} \phi_0^2 + \frac{\lambda_R^*}{32\pi^2} \left\{ \left[ m_{ph}^2 + \left(\xi_R - \frac{1}{6} \right) R \right] \ln \left( \frac{m_{ph}^2}{m_R^2} \right) - 2 F(m_{ph}^2,\{R\}) \right\}. \label{mph-eq}
\end{eqnarray}
This is the main result of this section. It shows that $m_{ph}^2$ can be completely expressed in terms of the renormalized parameters, showing a manifest invariance under changes of the regularization scale $\tilde{\mu}$. As a consequence both the equation for $\phi_0$ and the gap equation will also show these properties. Furthermore, it can be easily seen that in the free field limit, $\lambda_R \to 0$, we have $\lambda_R^* \to 0$ and consequently the physical mass becomes the renormalized mass, $m_{ph}^2 \to m_R^2$.

We close this section by defining the non-MS counterterms associated to the renormalized parameters in the following way
\begin{eqnarray}
 \delta \tilde{m} &=& m_B^2 - m_R^2 = \frac{m_R^2 \left[ 1 - \frac{\lambda}{32\pi^2} \ln \left( \frac{m_{R}^2}{\tilde{\mu}^2} \right) \right]}{1+\frac{\lambda}{16\pi^2\epsilon}} - m_R^2 = - \frac{\lambda_R^*}{32 \pi^2} \frac{m^2}{\left(1+\frac{\lambda}{16\pi^2\epsilon}\right)} \left[ \frac{2}{\epsilon} + \ln \left( \frac{m_{R}^2}{\tilde{\mu}^2} \right) \right] \notag \\
 &=& - m_B^2 \frac{\lambda_R^*}{32\pi^2} \left[ \frac{2}{\epsilon} + \ln \left( \frac{m_{R}^2}{\tilde{\mu}^2} \right) \right],
\end{eqnarray}
where we have used both Eqs. \eqref{mB} and \eqref{mR-mu} for $m_B^2$ and $m^2$, respectively, and then Eq. \eqref{m-lambda}. Similarly we obtain
\begin{eqnarray}
 \delta \tilde{\xi} &\equiv& \xi_B - \xi_R = - \left( \xi_B - \frac{1}{6} \right) \frac{\lambda_R}{32\pi^2} \left[ \frac{2}{\epsilon} + 1 + \ln \left( \frac{m_{R}^2}{\tilde{\mu}^2} \right) \right], \label{deltatildexi} \\
  \delta \tilde{\lambda} &\equiv& \lambda_{B2} - \lambda_R = - \lambda_B \frac{\lambda_R}{32\pi^2} \left[ \frac{2}{\epsilon} + 1 + \ln \left( \frac{m_{R}^2}{\tilde{\mu}^2} \right) \right].
\end{eqnarray}
These counterterms contain not only the poles in $\epsilon$ but also a $\tilde{\mu}$-dependent finite term. In this expressions the 1-loop limit can be easily taken by replacing $m_B^2 \to m_R^2$, $\xi_B \to \xi_R$ and $\lambda_{B2} \to \lambda_R$ in their right-hand-sides.

\section{Interacting fields in de Sitter spacetime}\label{DSSect}

In this section we apply the above results to de Sitter spacetimes. We consider the cosmological patch of de Sitter space in flat coordinates
\begin{equation}
 ds^2 = - dt^2 + e^{2Ht} d\vec{x}^2\,.
\end{equation}
The de Sitter spacetime is a solution to the vacuum Einstein equations with positive constant curvature, and 
as a consequence of its large degree of symmetry a possible solution to the mean field and gap equations
in this background is that with both $\phi_0$ and $\langle \varphi^2 \rangle = [G_1]/2$ constant.
In this geometry infrared divergences appear on $\langle \varphi^2 \rangle$ for massless minimally coupled free fields, while in
the case of interacting fields, the perturbative expansion is ill defined for light fields. 

\subsection{Physical mass equation}

We can study interacting fields in de Sitter spacetime in the Hartree approximation without much hustle by exploiting the fact that the gap Eq. \eqref{gap-eq} is actually the equation for the propagator of a free field with mass $m_{ph}^2$ (which is constant in de Sitter) and coupling to the curvature $\xi_R$. The solution of such an equation is known exactly in de Sitter for an arbitrary number of dimensions $n$. The most common form found in the literature for the coincidence limit $[G_1]$ is
\begin{equation}
 [G_1] = \frac{2 H^{n-2}}{(4\pi \mu^2)^{n/2}} \, \Gamma \left( 1 - \frac{n}{2} \right) \frac{\Gamma \left( \frac{n-1}{2} + \nu_n \right) \, \Gamma \left( \frac{n-1}{2} - \nu_n \right)}{\Gamma \left( \frac{1}{2} + \nu_n \right) \, \Gamma \left( \frac{1}{2} - \nu_n \right)},
\end{equation}
where $\nu_n^2 = \frac{(n-1)^2}{4} - \frac{m_{ph}^2}{H^2} - \xi_R n(n-1)$ and $R=n(n-1) H^2$. To make use of the results of Sec. II we can extract the function $F_{dS}(m_{ph}^2,R)$ from this expression, defined through the adiabatic expansion in Eq. \eqref{adiab-G1}. This is achieved by setting $n=4+\epsilon$ and expanding for $\epsilon \to 0$, holding $R$ fixed. From this comparison, detailed in Appendix C, we can extract the exact form of the function $F(m_{ph}^2,\{R\})$ for de Sitter spacetime,

\begin{eqnarray}
F_{dS}(m_{ph}^2,R) = R \, f(m_{ph}^2/R) = -\frac{R}{2} &\Biggl\{& \left(\frac{m_{ph}^2}{R} + \xi_R -\frac{1}{6} \right) \left[ \ln\left( \frac{R}{12 m_{ph}^2} \right) + g\left( m_{ph}^2/R + \xi_R \right) \right] \notag \\
&&- \left( \xi_R -\frac{1}{6} \right) - \frac{1}{18} \Biggr\}, \label{F-dS} 
\end{eqnarray}
with
\begin{equation}
g(y) \equiv \psi_{+} + \psi_{-} = \psi\left( \frac{3}{2} + \nu_4(y) \right) + \psi\left( \frac{3}{2} - \nu_4(y) \right), 
\label{little-g} 
\end{equation}
and where $R=12 H^2$, $\psi(x) = \Gamma^{'}(x)/\Gamma(x)$ is the DiGamma function and $\nu_4(y) = \sqrt{9/4-12y}$. The $F_{dS}$ function has all the expected properties, that is, it is written only in terms of renormalized parameters, it is independent of $\epsilon$ and $\tilde{\mu}$, and it satisfies the correct limits Eqs. \eqref{F-props}.

Then turning back to the physical mass equation, we can plug the expression for $F_{dS}$ in the general expression for the renormalized $m_{ph}^2$-equation \eqref{mph-eq}, to obtain
\begin{eqnarray}
  m_{ph}^2 + \xi_R R &=& m_R^2 + \xi_R R + \frac{\lambda_R^*}{2} \phi_0^2 + \frac{\lambda_R^*}{32 \pi^2} \left[ m_{ph}^2 + \left( \xi_R - \frac{1}{6} \right) R \right] \left[ \ln \left( \frac{R}{12 m_{R}^2} \right) + g\left( m_{ph}^2/R + \xi_R \right) \right] \nonumber\\
  &-& \frac{\lambda_R^*}{32 \pi^2} \left( \xi_R - \frac{1}{9} \right) R . \label{mph-eq-dS}
\end{eqnarray}
Note that in de Sitter this equation depends only on the combination 
\begin{equation}
\mathcal{M}_{ph}^2 \equiv m_{ph}^2 + \xi_R R\, . 
\end{equation}
Since we are interested in the infrared effects we consider $\mathcal{M}_{ph}^2 \ll R$ and expand the function $g(y)$ for $y = \mathcal{M}_{ph}^2/R \ll 1$, 
\begin{equation}
 g(y) \simeq - \frac{1}{4 y} + \frac{11}{6} - 2 \gamma_E + \frac{49}{9} y.
 \label{small-mass}
\end{equation}
This gives a pole in $\mathcal{M}_{ph}^2$ that can immediately be identified as the origin of the infrared divergence in $\langle \varphi^2 \rangle$. Hence, the solutions to the field and gap equations must be such that $\mathcal{M}_{ph}^2 > 0$. After expanding, multiplying by $y$ and keeping terms up to quadratic order in $y$ the gap equation \eqref{mph-eq-dS} reads
\begin{equation}
 A_{Mk} \, y^2 + \left[ B_{Mk} - \frac{\lambda_R \phi_0^2}{2R} \right]\, y + C_{Mk} = 0,
\end{equation}
where the coefficients are given by the following expressions,
\begin{subequations}
\begin{align}
A_{Mk} &=   1 - \frac{\lambda_R^*}{32\pi^2} \left[ a\left(R/12m_R^2 \right) - \frac{49}{54} \right], \\
B_{Mk} &= - \left( \frac{m_R^2}{R} + \xi_R \right) + \frac{\lambda_R^*}{32\pi^2} \left[ \frac{1}{6} a\left(R/12m_R^2 \right) + \xi_R + \frac{5}{36} \right], \nonumber \\
C_{Mk} &= - \frac{\lambda_R^*}{768 \pi^2},  
\end{align}
\label{CuadEq-dSren}
\end{subequations}
with
\begin{equation}
a(x) \equiv 11/6 - 2\gamma_E + \ln(x).
\label{a-definition}
\end{equation}
The $Mk$ subindex indicates we are using the Minkowski renormalization point. Note that  $C_{Mk}$ is always negative, while in principle $A_{Mk}$ and $B_{Mk}$ can have either sign. It can be seen that in order to have a real and positive solution $\mathcal{M}_{ph}^2 (\phi_0,R)$ for all $\phi_0$, it is necessary that $A_{Mk} > 0$, while $B_{Mk}$ is unconstrained. The solutions are
\begin{equation}
\mathcal{M}_{ph}^2 (\phi_0,R) = \frac{ -( R\, B_{Mk} - \frac{\lambda_R^* \phi_0^2}{2}) \pm \sqrt{ \left[ R B_{Mk} - \frac{\lambda_R^* \phi_0^2}{2} \right]^2 - 4 R^2\, A_{Mk} C_{Mk}  }  }{2 A_{Mk}}. 
\label{Mph-phi}
\end{equation}
Which branch is the appropriate one depends on the sign of $B_{Mk}$, but we can see that the solution is unique, as the other branch will be always negative.  For now we keep both branches. 

\subsection{Effective potential and symmetry breaking}\label{SBMink}

The effective potential is defined for fixed $R$ through its $\phi_0$-derivative that can be read from the field equation \eqref{field-eq} for constant $\phi_0$. This allows to find it by integration with respect to $\phi_0$,
\begin{equation}
 V_{eff}(\phi_0,R) = \int \left[ \mathcal{M}_{ph}^2(\phi_0^2,R) - \frac{1}{3} \lambda_R \phi_0^2 \right] \phi_0 \, d\phi_0 = \frac{1}{2} \int \mathcal{M}_{ph}^2(\phi_0^2,R) \, d\phi_0^2 - \frac{1}{12} \lambda_R \phi_0^4. 
\end{equation}
Using Eq. \eqref{Mph-phi} the result is 
\begin{eqnarray}
V_{eff}(\phi_0,R) &=& \mp \frac{2 R^2 \left(B_{Mk}-\frac{\lambda_R^* \phi_0^2}{2 R}\right)  \sqrt{\left(B_{Mk}-\frac{\lambda_R^* \phi_0^2}{2 R}\right)^2-4 A_{Mk} C_{Mk}}}{8 A_{Mk} \lambda_R^*} +  \frac{\frac{\lambda_R^*}{2} \phi_0^4 -2 B_{Mk} R \phi_0^2  }{8 A_{Mk}} \notag \\
&& \pm \frac{C_{Mk} R^2 \ln \left[\sqrt{\left(B_{Mk}-\frac{\lambda_R^* \phi_0^2}{2 R}\right)^2-4 A_{Mk} C_{Mk}}+B_{Mk}-\frac{\lambda_R^* \phi_0^2}{2 R}\right]}{\lambda_R^*} -\frac{\lambda_R \phi_0^4}{12}. \label{Veff-integrated}
\end{eqnarray}
This is a function of both $\phi_0$ and $R$ which is well defined for all $\phi_0$ as long as $A_{Mk} > 0$, as discussed above. 

The effective potential has an extreme at $\bar{\phi}_0 = 0$, which is the trivial solution to the field equation \eqref{field-eq} for constant $\phi_0$. It can be seen immediately that this must be a minimum, as the second derivative of $V_{eff}(\phi_0,R)$ at $\phi_0 = 0$ is $\mathcal{M}_{ph}^2(\phi_0=0,R)$, which must be positive in the Hartree approximation. We are interested in the question of whether there are other minima that break the $Z_2$ symmetry, that is, $\bar{\phi}_0 \neq 0$. According to equation \eqref{Veff}, this happens when
\begin{equation}
 \bar{\phi_0}^2 = \frac{3}{\lambda_R} \mathcal{M}_{ph}^2 (\bar{\phi_0},R).
\end{equation}
Using \eqref{Mph-phi} this condition can be recast as a quadratic equation for $\bar{\phi_0}^2$, whose solutions are
\begin{equation}
\bar{\phi_0}^2 =  \frac{3R}{\lambda_R} \left[ \frac{ - B_{Mk} \pm \sqrt{ B_{Mk}^2 - 4 \left(A_{Mk} - \frac{3\lambda_R^*}{2\lambda_R} \right) C_{Mk}  }  }{2 \left(A_{Mk} - \frac{3\lambda_R^*}{2\lambda_R} \right)} \right].
\label{bar-phi}
\end{equation}
As mentioned above, in the Hartree approximation the effective potential has always a minimum at $\bar{\phi}_0=0$. This implies that in order for another minimum to exist at $\bar{\phi}_0 \neq 0$, there must be also a maximum at smaller value of $\phi_0$, i.e. in between the two minima. For this reason we must consider both branches of \eqref{bar-phi}, and look for the conditions under which both solutions are real and positive. First of all, it can be seen that $A_{Mk} - 3\lambda_R^*/2\lambda_R < 0$ when $0<\lambda_R<1$, so the conditions are, in turn, $B_{Mk}^2 - 4 \left(A_{Mk} - 3\lambda_R^*/2\lambda_R \right) C_{Mk} > 0$ and $B_{Mk} > 0$. These conditions can be put together in the following one
\begin{equation}
 B_{Mk} - 2 \sqrt{ \left( \frac{3\lambda_R^*}{2\lambda_R} - A_{Mk} \right) |C_{Mk}| } > 0.
 \label{symm-breaking-cond}
\end{equation}
Note that imposing $B_{Mk} > 0$ selects the upper branch in \eqref{Mph-phi} and \eqref{Veff-integrated}.

\begin{figure}[h]
        \centering
        \begin{subfigure}[b]{0.5\textwidth}
                \centering
                \includegraphics[width=0.9\textwidth]{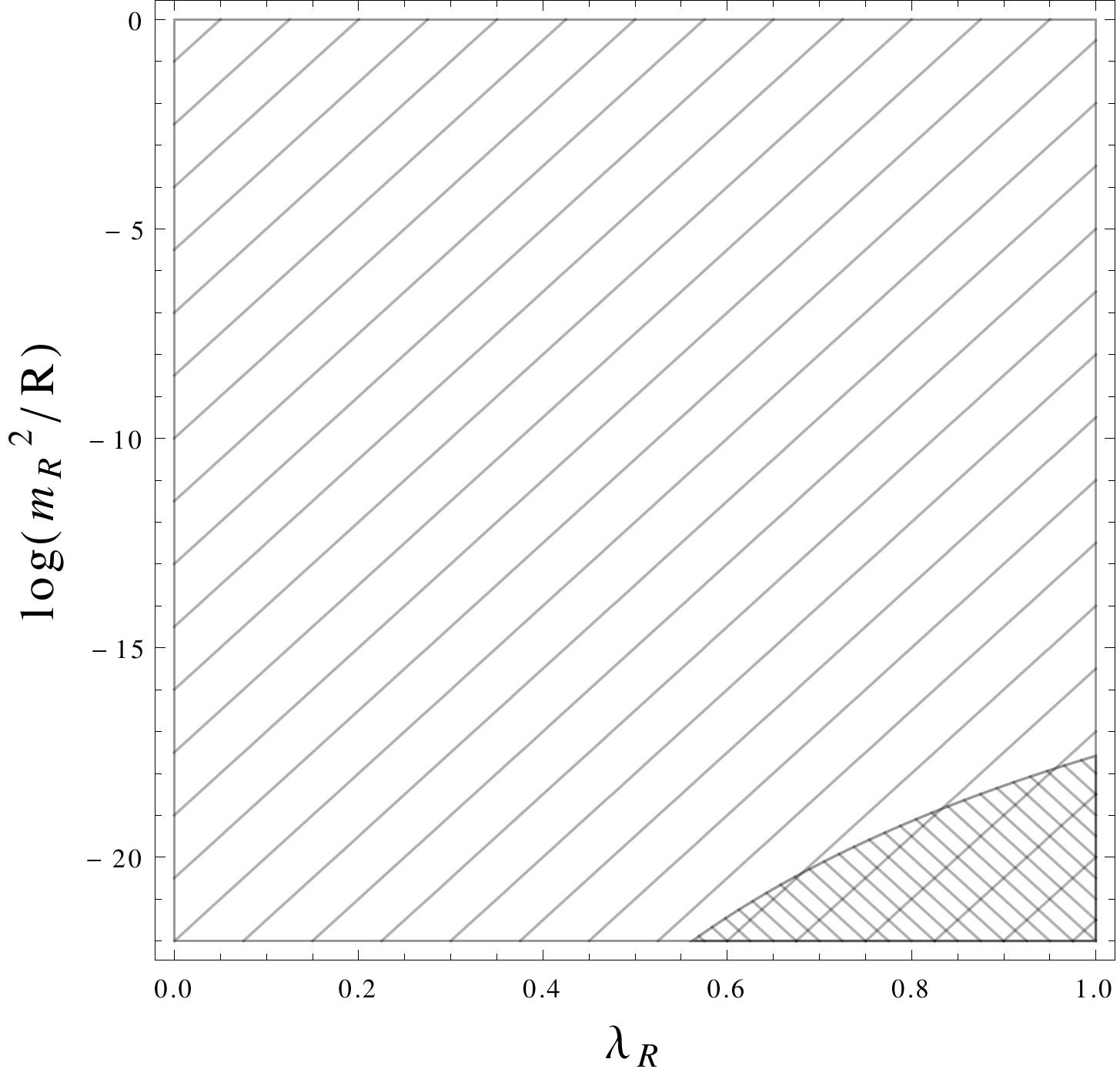}
                \caption{Minimally coupled case ($\xi_R = 0$)}
                \label{fig:RegionPlotMC-Mk}
        \end{subfigure}%
        \begin{subfigure}[b]{0.5\textwidth}
                \centering
                \includegraphics[width=0.9\textwidth]{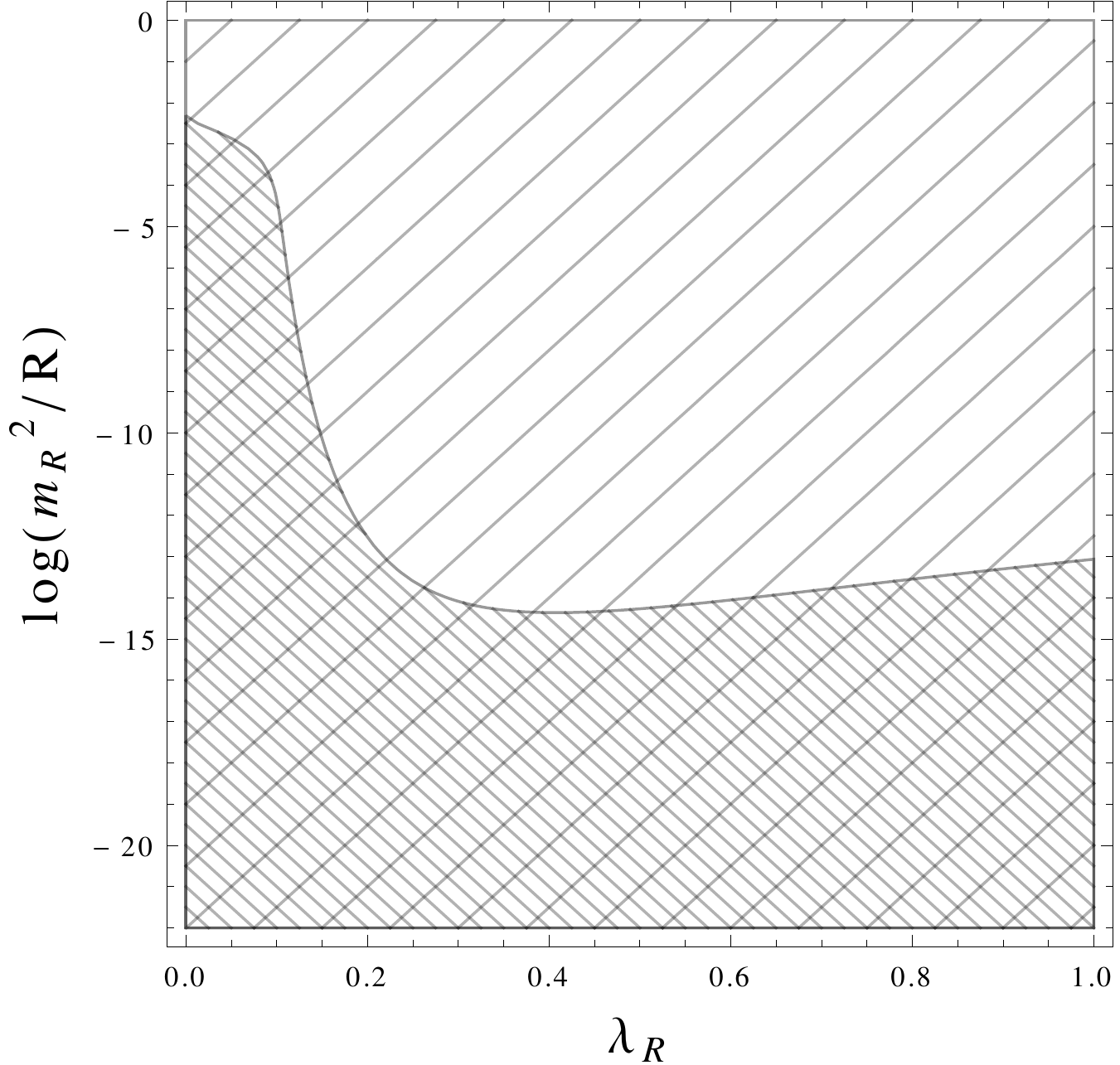}
                \caption{Non-Minimally coupled case ($\xi_R < 0$)}
                \label{fig:RegionPlotNMC-Mk}
        \end{subfigure}
        \caption{These plots show the regions in which the effective potential is well defined for all $\phi_0$ (low density stripes) and where the conditions for the existence of symmetry breaking solutions are met (high density stripes), as functions of $\lambda_R$ (horizontal axis) and $\log (m_R^2/R)$ (vertical axis), for $\xi_R = 0$ and $\xi_R = - 5\times 10^{-3}$. The first condition is met everywhere in both cases, while the spontaneous symmetry breaking exists for small $m_R^2/R$. The symmetry breaking region of the first plot moves further down and to the right when increasing $\xi_R$ to positive values.}\label{fig:RegionPlot-Mk}
\end{figure}

So now we can look for values of the parameters that satisfy simultaneously the conditions for having a well defined potential, $A_{Mk} > 0$, and for symmetry breaking \eqref{symm-breaking-cond}. The coefficients $A_{Mk}$, $B_{Mk}$ and $C_{Mk}$ depend on $m_R^2/R$, $\xi_R$ and $\lambda_R$. We plot in Fig. \ref{fig:RegionPlotNMC-Mk} the regions in the $\lambda_R$-$m_R^2/R$ plane for which each of these conditions hold, considering both the minimally coupled case $\xi_R = 0$, as well as $\xi_R < 0$. The $\xi_R > 0$ case is qualitatively similar to the minimally coupled case, but the region where symmetry breaking is possible shrinks.

\begin{figure}[h!]
       \centering
       \includegraphics[width=0.6\textwidth]{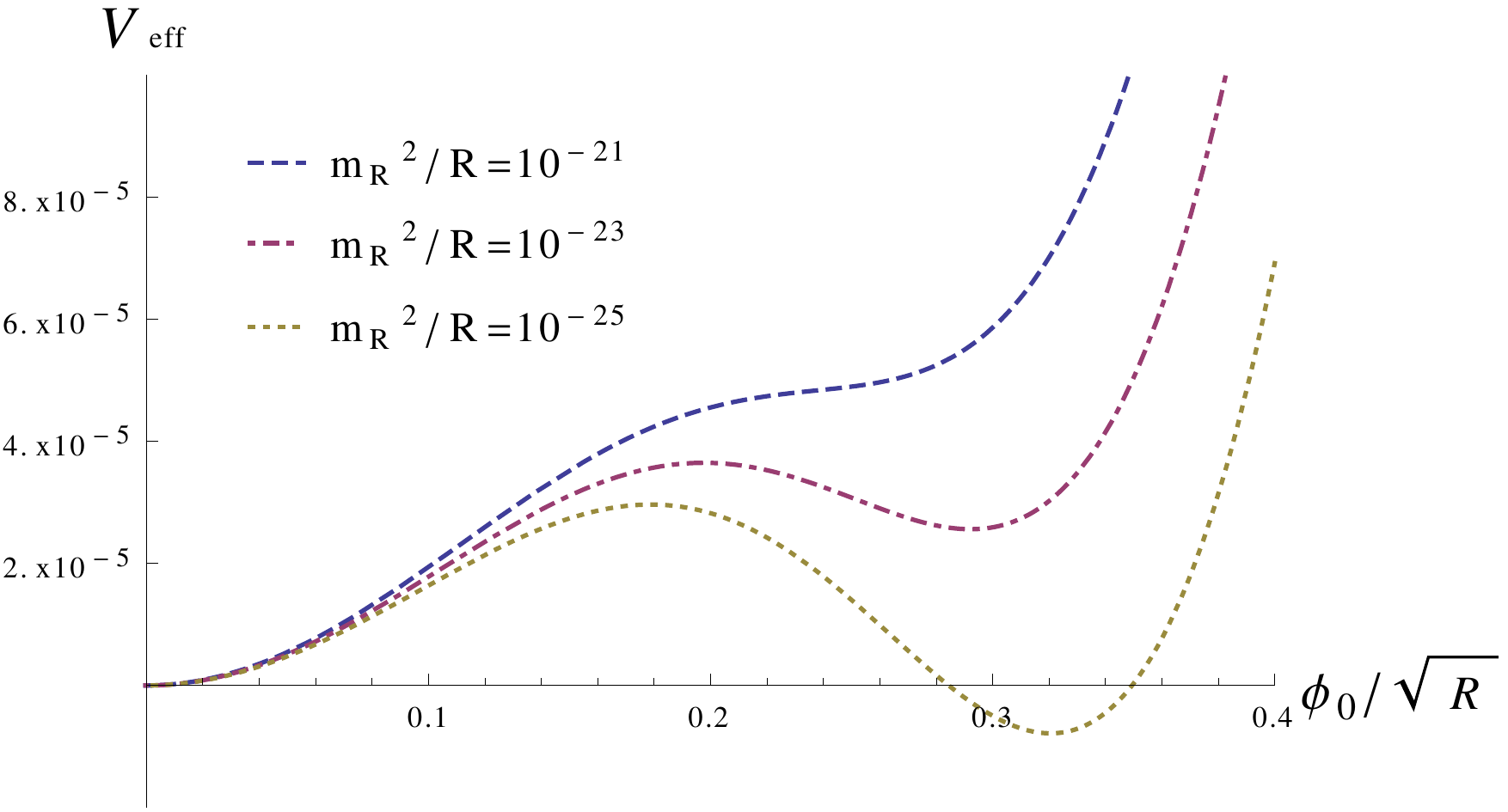}
       \caption{Effective Potential for different values of $m_R^2/R$. In all cases $\xi_R=0$ and $\lambda_R = 0.6$. There is always a minimum for $\phi_0 = 0$, while depending on the value of $m_R^2/R$ there can also be a minimum for $\phi_0 \neq 0$ with a maximum in between.}
       \label{fig:Veff}
\end{figure}

In Fig. \ref{fig:Veff} we show several curves of the effective potential for fixed parameters but varying values of $R$. It can be seen that the effective potential always has a minimum in $\bar{\phi_0} = 0$, while sometimes
 it can also have a second minimum for $\bar{\phi_0} \neq 0$.
We note that in order to see the symmetry breaking minimum it was necessary to consider a value of $m_R^2$ several orders of magnitude, more than 20, below $R$. 
Hence, this symmetry breaking minimum might be understood as  a peculiarity  of approaching  the massless limit. 
It is also worth mentioning  that this is not the usual kind of symmetry breaking/restoration scenario seen elsewhere in the literature where the symmetry is restored as $R$ increases, 
since in this case  the $R-$dependence  is just the opposite.

It is important to note that the reason why the $\bar{\phi}_0 \neq 0$ solutions are allowed is the presence of the $\lambda_R$ term in \eqref{field-eq}, which comes as a consequence of imposing the 2PI consistency relations. Otherwise it would happen as in the $1/N$ approximation or the usual Gaussian approach, where the absence of such term requires that for $\bar{\phi}_0 \neq 0$ we had $\mathcal{M}_{ph}^2 = 0$, and for that case there is no de Sitter invariant vacuum \cite{mazzi-paz,Serreau0}.

In the following section we will consider a more general choice of renormalization point, namely, de Sitter spacetime.

\section{de Sitter renormalization point}\label{RP}

In the previous sections we fixed the consistency conditions in Minkowski space by defining the renormalized parameters at $R=0$, and this was enough to show for example that the renormalized equations can be expressed in a $\tilde{\mu}$-independent way. However, it can be objected that
we are fixing the consistency conditions in a background different from the one we then want to study. As it was mentioned in Sec. III, it is not possible to impose that the consistency relations hold exactly for an arbitrary background, unless the spacetime has a constant curvature. Nevertheless, a de Sitter renormalization point with scalar curvature $R_0$ may feel more natural when the spacetime of interest is of FRW type, where one could fix the consistency conditions at given time by matching $R_0$ to the corresponding value of $R$ at that time. The consistency conditions could  then be incrementally violated as time progresses. Of course, even if we are studying the equations in de Sitter itself, although time-independent,  the matching   can be done exactly  only  if the variables  $R$ and $\phi_0$ are fixed beforehand. However, this is not the case we are considering here, as our purpose is to  analyze the dependence of the physical mass and effective potential on these two variables. Therefore,  the  natural  thing to do is to impose the consistency conditions    at a given renormalization point.
 Another point in favor of generalizing to de Sitter is that, as seen in the previous section, the Minkowski renormalization point did not allow for certain values of the MS-parameters that may be of interest, such as $m^2 < 0$. The aim of this section is to generalize the previous results to  the case where   the renormalization point is taken for a fixed de Sitter metric, applying it first to a general curved spacetime and then particularly to de Sitter itself.

We start by considering the consistency relations for the 4-point functions \eqref{4-pt-relation}, which gives a new definition of $\lambda_R$
\begin{equation}
 \frac{\delta^4 \Gamma_{1PI}[\phi_0]}{\delta \phi_1 \delta \phi_2 \delta \phi_3 \delta \phi_4} \Bigg|_{\phi_0=0,R=R_0} = - \lambda_R \delta_{12} \delta_{13} \delta_{14},
\end{equation}
where the notation $R=R_0$ implicitly implies we are evaluating in de Sitter spacetime, and also we avoided the use of a different notation for $\lambda_R$. The consistency relation for the 2-point functions Eq. \eqref{2-pt-relation} remains unchanged. Keeping this change in the definition in mind, the minimal subtraction renormalization proceeds in the same way as before and the bare couplings are found again to be given by Eqs. \eqref{bare-couplings}, and the finite gap equation in terms of the finite MS parameters is again  Eq. \eqref{phys-mass-fin}. The derivative of effective potential is read from the field equation as before, Eq. \eqref{Veff}, but now we change the definitions of the renormalized parameters by evaluating at $R=R_0$,
\begin{subequations}
\begin{align}
\mathcal{M}_R^2 &\equiv \frac{d^2 V_{eff}}{d \phi_0^2} \Bigg|_{\phi_0=0,R=R_0} = \mathcal{M}_{ph}^2(\phi_0=0,R=R_0), \label{ren-mass-def-dSren}  \\
\xi_R &\equiv \frac{d^3 V_{eff}}{d R \, d \phi_0^2} \Bigg|_{\phi_0=0,R=R_0} = \frac{d \mathcal{M}_{ph}^2}{d R} \Bigg|_{\phi_0=0,R=R_0}, \label{ren-xi-def-dSren} \\
\lambda_R &\equiv \frac{d^4 V_{eff}}{d \phi_0^4} \Bigg|_{\phi_0=0,R=R_0} = 3 \frac{d^2 \mathcal{M}_{ph}^2}{d \phi_0^2} \Bigg|_{\phi_0=0,R=R_0} - 2 \lambda_R. \label{ren-lambda-def-dSren}
\end{align}
\label{ren-quantities-dSren}
\end{subequations}
This redefinitions naturally lead to a generalization of Eqs. \eqref{ren-mu} relating the MS-parameters to the renormalized ones with an explicit dependence on $R_0$,
\begin{subequations}
\begin{align}
 m_R^2 &= \frac{m^2+\frac{\lambda}{16\pi^2}\left[ R_0 \frac{dF_{dS}}{dR}\Big|_{m_R^2,R_0}-F_{dS}(m_R^2,R_0) \right]}{\left[ 1 - \frac{\lambda}{32 \pi^2} \ln \left( \frac{m_R^2}{\tilde{\mu}^2} \right) \right]}, \label{mR-mu-dSren} \\
 \left( \xi_R - \frac{1}{6} \right) &= \frac{\left( \xi - \frac{1}{6} \right) - \frac{\lambda}{16\pi^2} \frac{dF_{dS}}{dR}\Big|_{m_R^2,R_0} }{ \left[1 - \frac{\lambda}{32\pi^2} - \frac{\lambda}{32\pi^2} \ln \left( \frac{m_R^2}{\tilde{\mu}^2} \right) \right] }, \label{xiR-mu-dSren} \\
  \lambda_R &= \frac{\lambda}{\left[ 1 - \frac{\lambda}{32\pi^2} - \frac{\lambda}{32\pi^2} \ln \left( \frac{m_{R}^2}{\tilde{\mu}^2} \right) - \frac{\lambda}{32\pi^2} \left( \frac{(\xi_R - \frac{1}{6}) R_0}{m_R^2} - 2 \frac{dF_{dS}}{d m_{ph}^2}\Big|_{m_R^2,R_0} \right) \right]}. \label{lambdaR-mu-dSren}
\end{align}
\label{ren-mu-dSren}
\end{subequations}
The original Minkowski spacetime equations are easily obtained by setting $R_0 \to 0$, which makes all the terms involving the $F_{dS}$ function 
or its derivatives to vanish according to its expected properties \eqref{F-props}. From now on we can use the explicit expression for $F_{dS}$ 
given in Eq. \eqref{F-dS}, and so
the previous expressions can be recast as
\begin{subequations}
\begin{align}
\mathcal{M}_R^2 - \frac{R_0}{6} &= \frac{m^2+(\xi-\frac{1}{6}) R_0 - \frac{\lambda R_0}{576 \pi^2}}{1 - \frac{\lambda}{32 \pi^2} \left[ \ln \left( \frac{R_0}{12 \tilde{\mu}^2} \right) + g(y_0) \right]}, \label{mR-mu-dSren2} \\
\left( \xi_R - \frac{1}{6} \right) &=  \frac{ \left( \xi - \frac{1}{6} \right) + \frac{\lambda}{32\pi^2} \left[ y_0 - \frac{1}{6} - \left(y_0 - \frac{1}{6} \right)^2 g'(y_0) - \frac{1}{18} \right] }{1 - \frac{\lambda}{32\pi^2} \left[ \ln \left( \frac{R_0}{12\tilde{\mu}^2} \right) + g(y_0) + \left(y_0 - \frac{1}{6} \right) g'(y_0) \right] }, \label{xiR-mu-dSren2} \\
 \lambda_R &= \frac{\lambda}{ 1 - \frac{\lambda}{32\pi^2} \left[ \ln \left( \frac{R_0}{12\tilde{\mu}^2} \right) + g(y_0) + \left( y_0 - \frac{1}{6} \right) g'(y_0) \right]}. \label{lambdaR-mu-dSren2}
\end{align}
\label{ren-mu-dSren2}
\end{subequations}
Here we are using the notation $y \equiv \mathcal{M}_{ph}^2/R$, and $y_0 = \mathcal{M}_{R}^2/R_0$. Also, Eqs. \eqref{mR-mu-dSren2} and \eqref{xiR-mu-dSren2} are not direct equivalents of Eqs. \eqref{mR-mu-dSren} and
 \eqref{xiR-mu-dSren}, but rather a mixing between them that will come in handy in what follows. 

Following the procedure outlined in Minkowski space, we can combine these equations in order to find $\tilde{\mu}$-independent relations between the MS-parameters and the renormalized parameters, namely
\begin{equation}
 \frac{m_B^2}{\lambda_{B2}} = \frac{m^2}{\lambda} = \frac{1}{\lambda_R} \left( \mathcal{M}_R^2 - \xi_R R_0 \right) + \frac{\left(\mathcal{M}_R^2 - \frac{R_0}{6} \right) }{32 \pi^2}.
 \label{m-lambda-dS}
\end{equation}
and
\begin{equation}
 \frac{\left( \xi_B - \frac{1}{6} \right)}{\lambda_B} = \frac{\left( \xi - \frac{1}{6} \right)}{\lambda} = \frac{\left( \xi_R - \frac{1}{6} \right)}{\lambda_R} + \frac{1}{32\pi^2} \left[ \left(y_0 - \frac{1}{6} \right)^2 g'(y_0) - \left(y_0 - \frac{1}{6} \right) + \frac{1}{18} \right].
 \label{xi-lambda-dS}
\end{equation}
These are as generalizations of Eqs. \eqref{m-lambda} and \eqref{xi-lambda} respectively. In Minkowski spacetime the first relation showed that the parameter $m^2$ could not be negative, while in de Sitter spacetime this relation is modified in such a way that makes that case possible, although not under general conditions.  More specifically, putting both of the above relations together we find that 
\begin{equation}
 \frac{1}{\lambda} \left( \frac{m^2}{R_0} + \xi \right) = \frac{y_0}{\lambda_R} + \frac{1}{32\pi^2} \left[ \left(y_0 - \frac{1}{6} \right)^2 g'(y_0) + \frac{1}{18} \right]
 +\frac{1}{6}\left(\frac{1}{\lambda}-\frac{1}{\lambda_R}\right).
\end{equation}
Then, considering that the Hartree approximation demands $y_0 > 0$, in which case it can be seen that $g'(y_0) > 0$, we conclude that the term in square
brackets in the right hand side should be positive definite, and hence the combination of MS-parameters given by $m^2 + \xi R_0$ must be positive
when $\lambda_R \geq \lambda$. We want to emphasize the importance of this result. The validity of the consistency conditions (which involve both finite and divergent parts) 
forbids to set $\lambda = \lambda_R$ and $m^2 + \xi R_0 < 0$ simultaneously, as was done in the literature when analysing spontaneous symmetry breaking in de Sitter space 
within the Hartree approximation \cite{Arai}. As a consequence of this result, one must allow for $\lambda_R < \lambda$, 
making it unavoidable to appeal 
to the effective potential in order to fix the finite part of the consistency condition Eq. \eqref{lambdas2-4}.

Carrying on with the renormalization in de Sitter space, we use Eqs. \eqref{ren-mu-dSren} to rewrite the gap Eq. \eqref{phys-mass-fin} in terms of the new renormalized 
parameters and $F_{dS}$,
\begin{eqnarray}
  m_{ph}^2 = m_R^2 + \frac{\lambda_R^*}{2} \phi_0^2 + \frac{\lambda_R^*}{32\pi^2} &\Biggl\{& \left[ m_{ph}^2 + \left(\xi_R - \frac{1}{6} \right) R \right] \ln \left( \frac{m_{ph}^2}{m_R^2} \right) \notag \\
  &&+ \left( m_{ph}^2 - m_R^2 \right) \left[ 2 \frac{dF_{dS}}{d m_{ph}^2}\Big|_{m_R^2,R_0} - \frac{(\xi_R - \frac{1}{6}) R_0}{m_R^2} \right] \label{mph-eq-dSren} \\ 
  &&+ 2\left[ F_{dS}(m_R^2,R_0) + \frac{dF_{dS}}{dR}\Big|_{m_R^2,R_0} \left( R - R_0 \right) - F(m_{ph}^2,\{R\}) \right] \Biggr\}. \notag
\end{eqnarray}

Setting the background metric $g_{\mu \nu}$ to de Sitter, that is, $F(m_{ph}^2,\{R\})=F_{dS}(m_{ph}^2,R)$, and using again Eq. \eqref{F-dS} and the variable $y = \mathcal{M}_{ph}^2/R = m_{ph}^2/R + \xi_R$, the gap equation reads
\begin{eqnarray}
y - \frac{1}{6} &=& \frac{R_0}{R} \left( y_0 - \frac{1}{6} \right) + \left( \xi_R - \frac{1}{6} \right) \left( 1 - \frac{R_0}{R} \right) + \frac{\lambda_R \phi_0^2}{2R} \notag \\
&+& \frac{\lambda_R}{32\pi^2} \Biggl\{ \left( y - \frac{1}{6} \right) \left[ \ln\left( \frac{R}{R_0} \right) + g(y) - g(y_0) - \left(y_0 - \frac{1}{6} \right) g'(y_0)  \right]  \notag\\
&-&  \left( 1 - \frac{R_0}{R} \right) \left(y_0 - \frac{1}{6} \right) + \left(y_0 - \frac{1}{6} \right)^2 g'(y_0) \Biggr\}.\label{gap-eq-dS} 
\end{eqnarray}
In order to solve it for $y(\phi_0,R)$ we consider the small mass case $\mathcal{M}_{ph}^2 \ll R$ ($y \ll 1$). The resulting quadratic equation now has the following coefficients
\begin{subequations}
\begin{align}
A_{dS} =&~ 1 - \frac{\lambda_R}{32\pi^2} \left[ a\left( \frac{R}{R_0} \right) - g(y_0) - \left(y_0 - \frac{1}{6} \right) g'(y_0) - \frac{49}{54} \right], \\
B_{dS} =& - \left[ \frac{R_0}{R} y_0 + \xi_R \left( 1 - \frac{R_0}{R} \right) \right]  \nonumber\\
 &+ \frac{\lambda_R}{32\pi^2} \left\{ \frac{1}{4} +  \frac{1}{6} \left[ a\left( \frac{R}{R_0} \right) - g(y_0) - \left(y_0 - \frac{1}{6} \right) g'(y_0) \right] \right.\nonumber \\
 &~~~~~~~~~~~~ + \left. \left( 1 - \frac{R_0}{R} \right) \left(y_0 - \frac{1}{6} \right) - \left(y_0 - \frac{1}{6} \right)^2 g'(y_0) \right\}, \\
C_{dS} =& -\frac{\lambda_R}{768\pi^2}, 
\end{align}
\label{CuadEq-dSren}
\end{subequations}
with $a(x)$ defined in Eq. \eqref{a-definition}.
The analysis of this equation and its solutions now proceeds exactly the same as in Minkowski spacetime, but now there is one new parameter that must be taking into account, $R_0$.

\subsection{Spontaneous symmetry breaking}

\begin{figure}[h]
        \centering
	\includegraphics[width=0.45\textwidth]{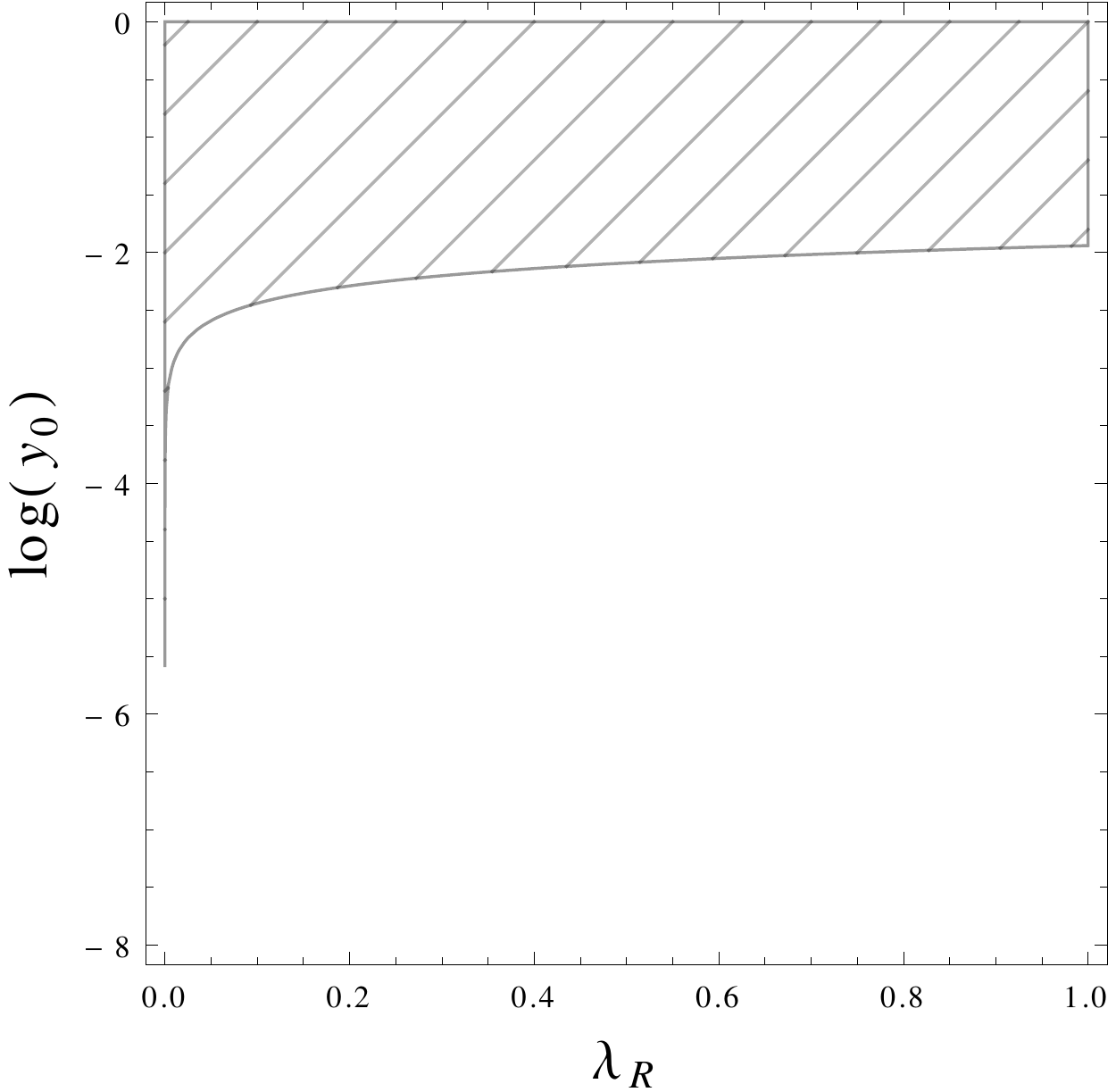}
        \caption{This plot shows, for $R=R_0$, the region in which the effective potential is well defined for all $\phi_0$ (low density stripes) in the $\lambda_R$-$\log(y_0)$ plane. There is no region in which the symmetry breaking conditions are met. There is no other free parameter.}
        \label{fig:RegionPlot-dS-R0}
\end{figure}

Let us first consider $R=R_0$. The first interesting fact is that the gap equation \eqref{gap-eq-dS} does not depend explictly on $\xi_R$ when $R=R_0$, but rather only through the combination $y_0 = m_R^2/R + \xi_R$. This simplifies the analysis by leaving only two parameters, $y_0$ and $\lambda_R$. In Fig. \ref{fig:RegionPlot-dS-R0} we show in the $\lambda_R$-$\log(y_0)$ plane that while the effective potential can be well defined in some region (low density stripes), the symmetry breaking conditions are not met. We emphasize that there is no other free parameter when $R=R_0$, as all the combinations of $m_R^2/R_0$ and $\xi_R$ such that $y_0 > 0$ are accounted for.

\begin{figure}[h!]
        \centering
        \begin{subfigure}[b]{0.5\textwidth}
                \centering
                \includegraphics[width=0.9\textwidth]{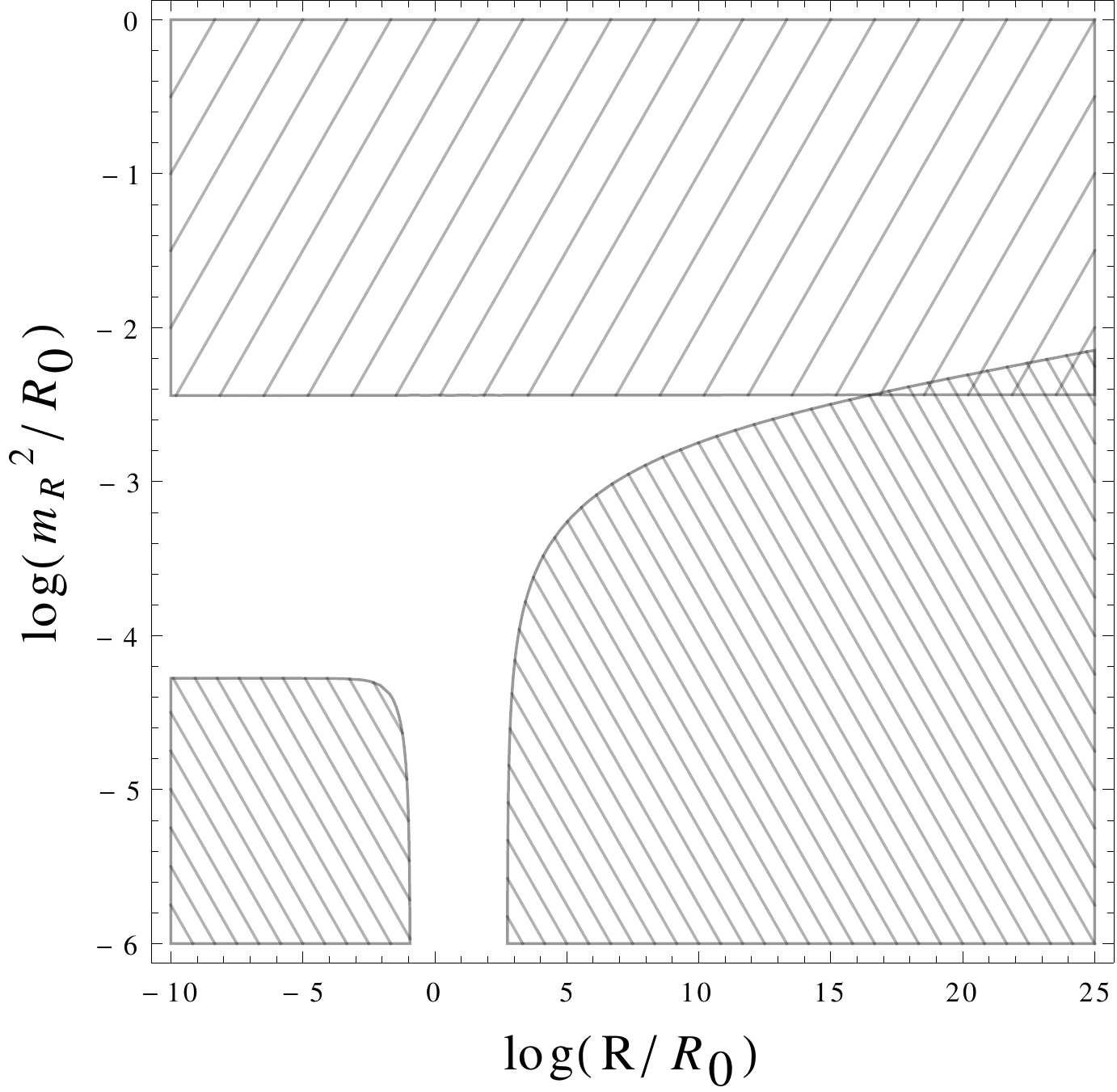}
                \caption{Minimally coupled case ($\xi_R = 0$)}
                \label{fig:RegionPlotMC-dS}
        \end{subfigure}%
        \begin{subfigure}[b]{0.5\textwidth}
                \centering
                \includegraphics[width=0.9\textwidth]{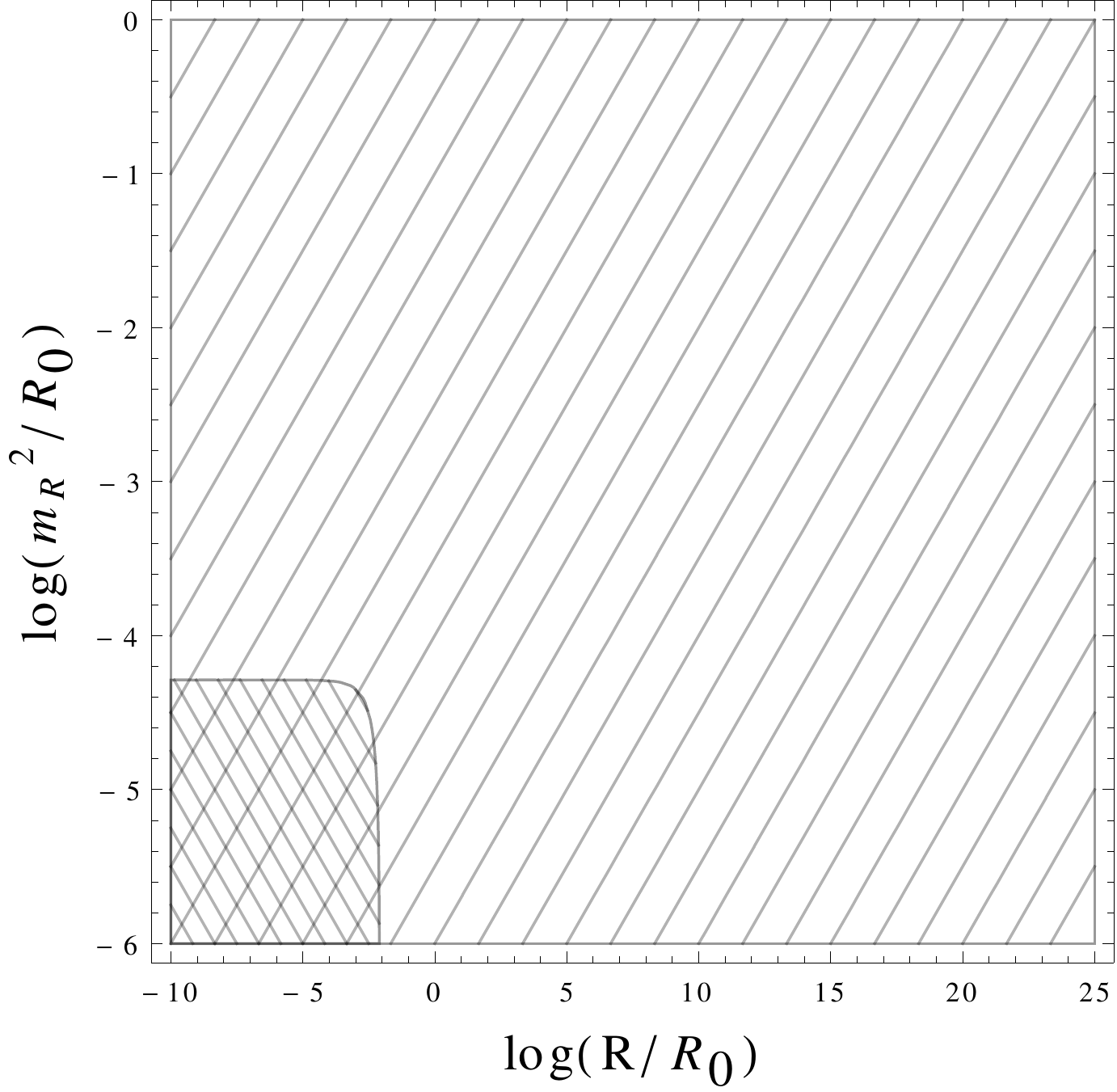}
                \caption{Non-Minimally coupled case ($\xi_R > 0$)}
                \label{fig:RegionPlotNMC-dS}
        \end{subfigure}
        \caption{These plots show the regions in which the effective potential is well defined for all $\phi_0$ (low density stripes) and where the conditions for the existence of symmetry breaking solutions are met (high density stripes), as functions of $\log (R/R_0)$ (horizontal axis) and $\log (m_R^2/R_0)$ (vertical axis), for $\lambda_R = 0.1$. For $\xi_R = 0$ both regions only overlap at $R \gg R_0$ and with $m_R^2/R_0$ in a certain range, while for $\xi_R = 4 \times 10^{-3}$ the effective potential is well defined everywhere and spontaneous symmetry breaking exists for small $R/R_0$ and $m_R^2/R_0$. The plots do not change qualitatively by varying $\lambda_R$.}\label{fig:RegionPlot-dS}
\end{figure}

\begin{figure}[h!]
    \includegraphics[width=0.6\textwidth]{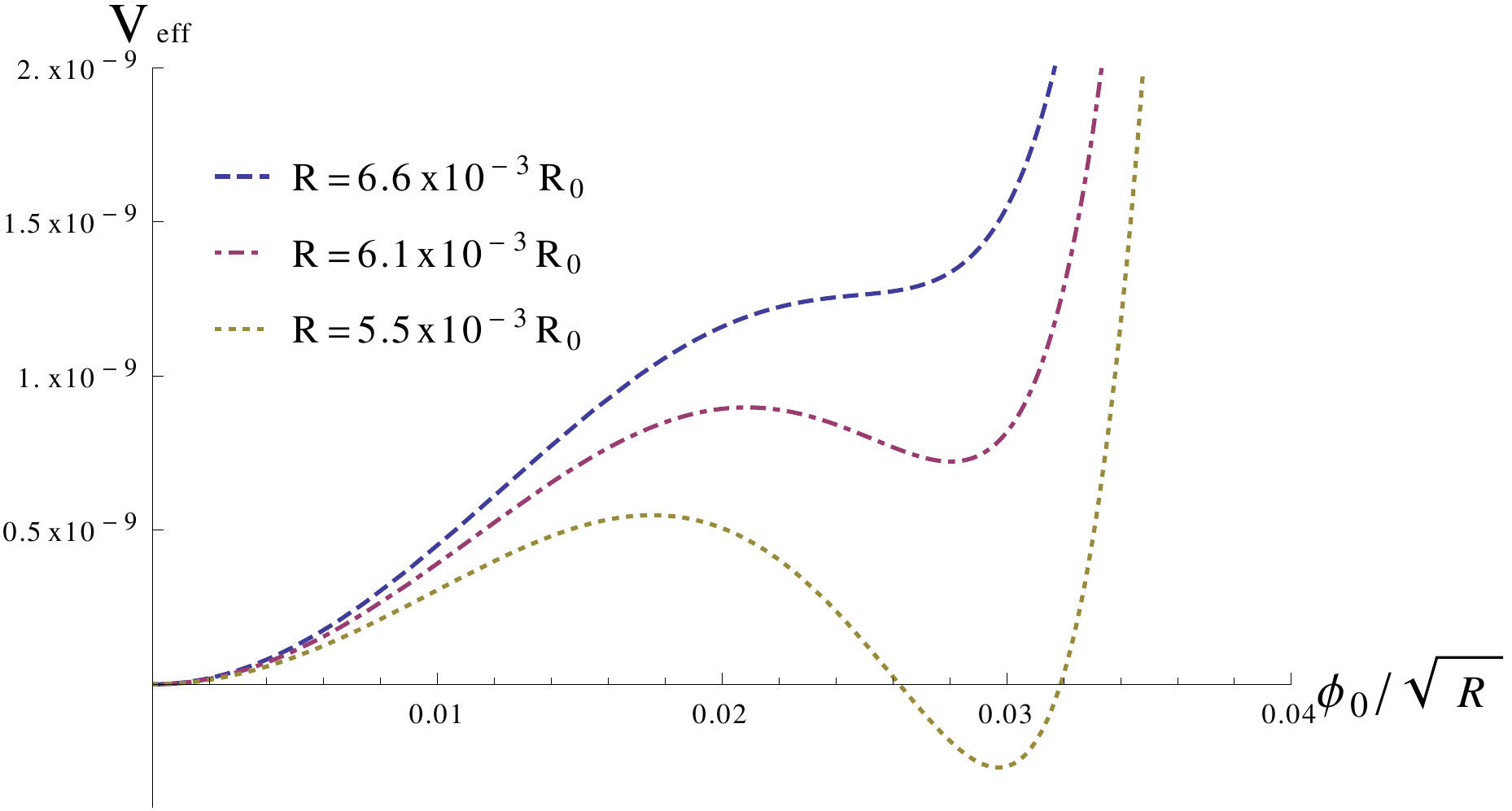}
    \caption{Effective potential for $m_R^2/R_0 = 10^{-5}$, $\xi_R = 4\times 10^{-3}$ and $\lambda_R = 0.1$, for different values of $R/R_0$. Symmetry
    breaking is exhibited for small $R/R_0$, while for larger $R/R_0$ the symmetry is restored. The critical value in this case is $R/R_0 \sim 6.5\times 10^{-3}$.}
    \label{fig:potdS}
\end{figure}

Finally, we consider the case $R \neq R_0$. We show in Fig. \ref{fig:RegionPlot-dS} plots in the $\log (R/R_0)$-$\log (m_R^2/R_0)$ plane of the regions of interest for fixed $\xi_R$ and $\lambda_R = 0.1$. Note that these plots are in a different plane of the parameter space in contrast to the previous plots we analysed. In the minimally coupled case, both regions overlap only at $R \gg R_0$ and with $m_R^2/R_0$ in a certain range. In particular this implies that in the low mass limit $m_R^2 \ll R_0,R$ there is no possible symmetry breaking. Moreover, the effective potential is not even well defined for all $\phi_0$. Note that this is different to what we have seen in Fig. \ref{fig:RegionPlotMC-Mk} in Minkowski space, where the situation was $R_0 < m_R^2 \ll R$, that is, the flat limit was taken first. It is well known that the flat and massless limits in de Sitter do not commute.
If we now allow for $\xi_R \neq 0$, we find a different situation. Both regions can overlap for certain values of the parameters, allowing for a well defined effective potential with symmetry breaking. Some examples of such a potential are shown in Fig. \ref{fig:potdS} for different values of $R/R_0$. Symmetry breaking occurs for small $R/R_0$ and then the symmetry is restored for larger $R/R_0$.

 \newpage

\section{Conclusions}\label{ConcluSect} 

In this section we will summarize the results obtained in this paper and will discuss the relation with previous works on the subject.

We have considered a single, selfinteracting scalar field with $Z_2$ symmetry in curved spacetimes, using a nonperturbative approach based on the 2PI EA. 
In the lowest approximation (a local approximation of the 2PI EA) the formalism reproduces the usual Hartree (or Gaussian) approximation, that can also be derived using a variational approach or a resummation of a particular class of Feynman diagrams. However, when considered in the context of the 2PI EA, there are some consistency conditions which, although automatically satisfied in the full theory, are  not fulfilled for particular approximations (this is the case when the approximation is not a systematic expansion in powers of
a small parameter).  This fact is well known in flat spacetime, and the consistency conditions can be forced by allowing for more than one counterterm for
each mass or coupling constant of the theory \cite{Bergesetal}. Our first goal in this paper has been to  show that this ``consistent renormalization procedure" 
can be applied to the mean field and gap equations in  general curved spacetimes.
Several explicit calculations previously performed in flat spacetime used as regulator  a momentum cutoff (see for instance \cite{Marko}). 
This is problematic in curved backgrounds, and in order to maintain the covariance of the regularized theory we used dimensional regularization.  

The consistent renormalization procedure has been partially extended to curved spacetimes in some recent works \cite{Arai}, where the renormalization 
of the mean field and gap equations has been analyzed. Our results in Sec. III are more complete than
those in Ref.\cite{Arai}. We have included the full adiabatic expansion of the propagator, we have written 
the renormalized equations in terms of the physical parameters defined from the effective potential, and we have shown explicitly that, 
when expressed in terms of these physical parameters, the equations are independent of the regularization scale $\tilde\mu$ introduced by dimensional
regularization.   


In the last part of the paper (Secs. IV and V), we have focused on de Sitter spacetime. We have written the explicit form of  the mean field and 
gap  equations for this particular metric, and we have computed  the effective potential. For this, we have considered the renormalized parameters at $\phi_0=0$ as defined from the effective potential corresponding, in turn, to Minkowski spacetime and de Sitter spacetime with a given curvature $R_0$. Then, we performed an  analysis of the effective potential for different values of 
 the curvature of the original de Sitter spacetime $R$. This analysis  is somehow  analogous to the one  done in \cite{Reinosa} for a self-interacting scalar field at finite temperature, where the renormalization point is 
 chosen for a fixed finite value of the temperature. 
 
A relevant conclusion of our work is that the Hartree approximation and the application of the consistency conditions 
 impose restrictions on the 
 MS- parameters of the theory, as discussed in Secs IV and V. On the one hand, the renormalized mass $m_R^2$ must be
 positive. On the other hand, the definition of the renormalized $\lambda_R$ in terms of the fourth derivative of the effective potential and 
 the consistency condition for the 4-point function imply $\tilde\mu$-independent relations between MS and renormalized parameters. From them,
 it is clear that some choices of the MS-parameters may not be compatible with the consistency conditions. In particular, one cannot take
 $\lambda=\lambda_R$, $\xi=0$,  and $m^2<0$ simultaneously. 
 
Our results show that the $Z_2$ symmetry can be spontaneously broken when one uses the consistent renormalization, although not under general conditions.
This is to be contrasted with previous results obtained using a standard renormalization of the theory, as can be easily seen from 
the gap and mean value equations. Indeed,
when the consistency relations are not taken into account and the presence of different counterterms is not allowed, Eqs. (\ref{field-eq}) and (\ref{gap-eq}) become \cite{mazzi-paz}
\begin{eqnarray}
 \left( -\square + m_{ph}^2 + \xi_R R \right) \phi_0(x) &=& 0, \label{field-eq-usual} \\
 \left( -\square + m_{ph}^2 + \xi_R R \right) G_1(x,x') &=& 0.\label{gap-eq-usual}\, 
\end{eqnarray}
(note that the extra term in Eq. (\ref{field-eq}) comes from the consistency conditions). If there were spontaneous symmetry breaking in de Sitter spacetime, one would have
a solution to these equations with a constant and non-vanishing $\phi_0$.  If this were the case, one should have  
\begin{equation}
\square G_1(x,x') = 0.\label{gap-eq-usual-massless}
\end{equation}
However, it is well known that there is no de Sitter invariant propagator for a minimally coupled and massless scalar field, and therefore a constant
$\phi_0\neq 0$ cannot be a solution of Eqs. (\ref{field-eq-usual}, \ref{gap-eq-usual}). Thus, there is no spontaneous symmetry breaking when one uses the standard renormalization.

A similar situation holds in the large $N$ limit of $O(N)$ theories. In this case, the consistency relations are satisfied automatically order by order in powers of $1/N$, and there is no need for additional counterterms. The mean field and gap equations are once more given by Eqs. (\ref{field-eq-usual}) and (\ref{gap-eq-usual}) \cite{mazzi-paz},
and the same argument applies \cite{Serreau0,mazzi-paz}.

The results of this paper are not conclusive about the occurrence of spontaneous symmetry breaking of the $Z_2$ symmetry. It is not clear whether 
the existence of solutions with $\phi_0\neq 0$ is an artifact of the Hartree approximation or not.   It is plausible that the inclusion of the setting 
sun diagram in the computation of
the 2PI EA restores the $Z_2$ symmetry. We hope to address this interesting issue in a forthcoming work. However, there are several technical complications
to be sorted. On the one hand, the use of the CTP formalism will be unavoidable when considering nonlocal terms in the 2PI EA. On the other hand, the inclusion of
higher loops in the 2PI EA induces some subtle points in the renormalization even in flat spacetime \cite{Marko}, that will have their counterpart in curved spaces.

Finally, it is important to show that this consistent renormalization procedure can be extended for general curved spacetimes,
to make finite not only the mean field and gap equations of the matter sector of the theory, but also the gravitational sector. 
This is the main goal of the forthcoming paper II \cite{Nos2}.

\section*{Acknowledgements}
This research was supported in part by ANPCyT, CONICET and UBA. F.D.M and L.G.T would like to thank the hospitality of ICTP,
where part of this work has been done.  We would like to thank Guillem P\'erez Nadal for useful discussions at the initial stages of this work.

\newpage

\section*{Appendix A: Consistency relations in the 2PI formalism }
\label{apB}


In this section we briefly review the derivation of the consistency conditions we used to establish a relation between the different counterterms, which are Eqns.. (\ref{2-pt-relation}) and (\ref{4-pt-relation}). 
Let us start by recalling that the 1PI resummed effective action is obtained by evaluating the 2PI EA in the solution $\bar{G}(\phi_0)$  of the gap equation,  
\begin{equation}
 \Gamma_{1PI}[\phi_0] = \Gamma_{2PI}[\phi_0,\bar{G}(\phi_0)].
 \label{gamma-1PI}
\end{equation}
From \eqref{gamma-2PI-alt} the gap equation is, formally,
\begin{equation}\label{gaux}
\frac{\delta \Gamma_{2PI}}{\delta G_{12}} \Bigg|_{\bar{G}} = - \frac{i}{2} \bar{G}^{-1}_{12} + \frac{i}{2} G^{-1}_{0,12} + \frac{\delta \Gamma_{int}}{\delta G_{12}} \Bigg|_{\bar{G}} = 0,
\end{equation}
or, equivalently, 
\begin{equation}
 \bar{G}^{-1}_{12}(\phi_0) = G^{-1}_{0,12} - \bar{\Sigma}_{12}(\phi_0),
 \label{gap-eq-c}
\end{equation} using the definition of the self-energy
\begin{equation}
 \bar{\Sigma}_{12}(\phi_0) \equiv 2i \frac{\delta \Gamma_{int}}{\delta G_{12}} \Bigg|_{\bar{G}}.
\end{equation}
The first derivative of $ \Gamma_{1PI}[\phi_0]$ with respect to $\phi_0$,  which equated to zero gives the field equation,   can be written as
\begin{eqnarray}
 \Gamma_1^{(1)} &=& \frac{\delta \Gamma[\phi_0]}{\delta \phi_1} \Bigg|_{\bar{\phi_0}} = \frac{\delta \Gamma_{2PI}}{\delta \phi_1} + \frac{\delta \Gamma_{2PI}}{\delta G_{ab}} \bigg|_{\bar{G}} \, \frac{\delta \bar{G}_{ab}}{\delta \phi_1} = \frac{\delta \Gamma_{2PI}}{\delta \phi_1} \nonumber\\
		&=& i G_{0,1a}^{-1} \phi_a + \frac{\delta \Gamma_{int}}{\delta \phi_1} \Bigg|_{\bar{G}} ,
\label{1-pt} 
\end{eqnarray} where in the third equality we used Eq. (\ref{gaux}).

As mentioned in the main text, there are  multiple definitions of
a given $n$-point function. We focus here on the two- and four-point functions.

On one side, we have the   two- and four-point functions that can be obtained by taking functional derivatives of  $\Gamma_{1PI}[\phi_0]$ with respect to $\phi_0$:
\begin{equation}
 \Gamma_{12}^{(2)} = \frac{\delta^2 \Gamma[\phi_0]}{\delta \phi_1 \delta \phi_2} \Bigg|_{\bar{\phi_0}},
\end{equation}
\begin{equation}
 \Gamma_{1234}^{(4)} = \frac{\delta^4 \Gamma[\phi_0]}{\delta \phi_1 \delta \phi_2 \delta \phi_3 \delta \phi_4} \Bigg|_{\bar{\phi_0}}.
\end{equation}
Functionally deriving Eq. \eqref{1-pt}, we obtain the  following expression for the 2-point function
\begin{equation}
 \frac{\delta^2 \Gamma[\phi_0]}{\delta \phi_1 \delta \phi_2} \Bigg|_{\bar{\phi_0}} = i G_{0,12}^{-1} + \frac{\delta^2 \Gamma_{int}}{\delta \phi_1 \delta \phi_2} \Bigg|_{\bar{G}} + \frac{\delta^2 \Gamma_{int}}{\delta \phi_1 \delta G_{ab}} \Bigg|_{\bar{G}} \frac{\delta \bar{G}_{ab}}{\delta \phi_2}.
\end{equation}
Using that
\begin{equation}
\frac{\delta \bar{G}_{12}}{\delta \phi_3} = \frac{\delta}{\delta \phi_3} \left( \bar{G}_{1a} \bar{G}^{-1}_{ab} \bar{G}_{b2} \right) = \frac{\delta \bar{G}_{12}}{\delta \phi_3} + \frac{\delta \bar{G}_{12}}{\delta \phi_3} + \bar{G}_{1a} \bar{G}_{b2} \frac{\delta \bar{G}^{-1}_{ab}}{\delta \phi_3},
\end{equation}
and that from Eq. \eqref{gap-eq-c} we have
\begin{equation}
\frac{\delta \bar{G}_{12}}{\delta \phi_3} = \bar{G}_{1a} \bar{G}_{b2} \frac{\delta \bar{\Sigma}_{ab}}{\delta \phi_3},
\label{Gphi}
\end{equation}
we can write 
\begin{equation}
 \frac{\delta^2 \Gamma[\phi_0]}{\delta \phi_1 \delta \phi_2} \Bigg|_{\bar{\phi_0}} = i G_{0,12}^{-1} + \frac{\delta^2 \Gamma_{int}}{\delta \phi_1 \delta \phi_2} \Bigg|_{\bar{G}} + \frac{\delta^2 \Gamma_{int}}{\delta \phi_1 \delta G_{ab}} \Bigg|_{\bar{G}} \bar{G}_{ac} \bar{G}_{bd} \frac{\delta \bar{\Sigma}_{cd}}{\delta \phi_2}.
\label{2-pt}
\end{equation}

Moreover, one can show that
\begin{eqnarray}
 \frac{\delta \bar{\Sigma}_{12}}{\delta \phi_3} &=& 2i \frac{\delta^2 \Gamma_{int}}{\delta \phi_3 \delta G_{12}} \Bigg|_{\bar{G}} + 2i \frac{\delta^2 \Gamma_{int}}{\delta G_{12} \delta G_{ab}} \Bigg|_{\bar{G}} \frac{\bar{G}_{ab}}{\delta \phi_3} \\
 &=& 2i \frac{\delta^2 \Gamma_{int}}{\delta \phi_3 \delta G_{12}} \Bigg|_{\bar{G}} + \frac{i}{2} \bar{\Lambda}_{12,ab} \bar{G}_{ac} \bar{G}_{bd} \frac{\delta \bar{\Sigma}_{cd}}{\delta \phi_3},
\label{eq-sigma1-autoc}
\end{eqnarray}
where $\bar{\Lambda}_{12,34}$ is defined by
\begin{equation}
 \bar{\Lambda}_{12,34} \equiv 4 \, \frac{\delta^2 \Gamma_{int}}{\delta G_{12} \delta G_{34}} \Bigg|_{\bar{G}}.
 \label{barLambda}
\end{equation}
Hence, $\delta \bar{\Sigma}_{12}/\delta \phi_3$ satisfies a self-consistent equation. In a similar way, one can define a 4-point
vertex function as the self-consistent solution of the following
equation
\begin{equation}
 \bar{V}_{12,34} = \bar{\Lambda}_{12,34} + \frac{i}{2} \bar{\Lambda}_{12,ab} \bar{G}_{ac} \bar{G}_{bd} \bar{V}_{cd,34},
\label{Vert}
\end{equation}
which in matrix form reads
\begin{equation}
 \bar{V} = \bar{\Lambda} + \frac{i}{2} \bar{\Lambda} \bar{G}^2 \bar{V} = \bar{\Lambda} + \frac{i}{2} \bar{V} \bar{G}^2 \bar{\Lambda},
\end{equation}
where the last equality follows from the symmetry properties of  $\bar{V}$ (which are the same as $\Gamma^{(4)}_{12,34}$) and $\bar{\Lambda}$.

In order to obtain the  4-point function one should take two more derivatives, however, this would become very cumbersome. Therefore, let us restrict ourselves to
the case of $Z_2$ symmetric theories where a  simplification occurs. In this case, $n$-point functions with odd n vanish at $\phi_0=0$, for instance
\begin{equation}
 \frac{\delta \bar{\Sigma}_{12}}{\delta \phi_3} \Bigg|_{\phi_0 = 0} = 2i \frac{\delta^2 \Gamma_{int}}{\delta \phi_3 \delta G_{12}} \Bigg|_{\phi_0 = 0} = 0.
\end{equation}
Taking this into account, the 4-point function results 
\begin{eqnarray}
 \Gamma_{1234}^{(4)} &=& \frac{\delta^4 \Gamma_{int}}{\delta \phi_1 \delta \phi_2 \delta \phi_3 \delta \phi_4} \Bigg|_{\bar{G}} + \frac{\delta^3 \Gamma_{int}}{\delta \phi_1 \delta \phi_2 \delta G_{ab}} \Bigg|_{\bar{G}} \bar{G}_{ac} \bar{G}_{db} \frac{\delta^2 \bar{\Sigma}_{cd}}{\delta \phi_3 \delta \phi_4} \notag \\
&+& \frac{\delta^3 \Gamma_{int}}{\delta \phi_1 \delta \phi_3 \delta G_{ab}} \Bigg|_{\bar{G}} \bar{G}_{ac} \bar{G}_{db} \frac{\delta^2 \bar{\Sigma}_{cd}}{\delta \phi_2 \delta \phi_4} + \frac{\delta^3 \Gamma_{int}}{\delta \phi_1 \delta \phi_4 \delta G_{ab}} \Bigg|_{\bar{G}} \bar{G}_{ac} \bar{G}_{db} \frac{\delta^2 \bar{\Sigma}_{cd}}{\delta \phi_2 \delta \phi_3} .
\end{eqnarray}
From Eq. \eqref{eq-sigma1-autoc}, using the  $Z_2$ symmetry, we obtain a self-consistent equation
\begin{eqnarray}
 \frac{\delta^2 \bar{\Sigma}_{12}}{\delta \phi_3 \delta \phi_4} &=& 2i \frac{\delta^3 \Gamma_{int}}{\delta \phi_3 \delta \phi_4 \delta G_{12}} \Bigg|_{\bar{G}} + \frac{i}{2} \bar{\Lambda}_{12,ab} \bar{G}_{ac} \bar{G}_{bd} \frac{\delta^2 \bar{\Sigma}_{cd}}{\delta \phi_3 \delta \phi_4}
\label{eq-sigma2-autoc}
\end{eqnarray}
whose solution can be expressed in terms of the function $\bar{V}_{12,34}$, solution of \eqref{Vert},
\begin{eqnarray}
 \frac{\delta^2 \bar{\Sigma}_{12}}{\delta \phi_3 \delta \phi_4} &=& 2i \frac{\delta^3 \Gamma_{int}}{\delta \phi_3 \delta \phi_4 \delta G_{12}} \Bigg|_{\bar{G}} + \frac{i}{2} \bar{V}_{12,ab} \bar{G}_{ac} \bar{G}_{bd} 2i \frac{\delta^3 \Gamma_{int}}{\delta \phi_3 \delta \phi_4 \delta G_{cd}} \Bigg|_{\bar{G}} \notag \\
 &=& i \left( \Lambda_{34,12} + \frac{i}{2} \bar{V}_{12,ab} \bar{G}_{ac} \bar{G}_{bd} \Lambda_{34,cd} \right)
\end{eqnarray}
where we used the definition
\begin{equation}
 \Lambda_{12,34} \equiv 2 \frac{\delta^3 \Gamma_{int}}{\delta \phi_1 \delta \phi_2 \delta G_{34}} \Bigg|_{\bar{G}}.
\end{equation}
Then, the  4-point function can be written as
\begin{eqnarray}
 \Gamma_{1234}^{(4)} = \frac{\delta^4 \Gamma_{int}}{\delta \phi_1 \delta \phi_2 \delta \phi_3 \delta \phi_4} \Bigg|_{\bar{G}} + \frac{i}{2} \left[ \Lambda_{12,ab} \bar{G}_{ac} \bar{G}_{db} \Lambda_{cd,34} + \frac{i}{2} \Lambda_{12,ab} \bar{G}_{ac} \bar{G}_{db} \bar{V}_{cd,ef} \bar{G}_{eh} \bar{G}_{if} \Lambda_{hi,34} + perm \right],\nonumber
\end{eqnarray}
or, in matrix form
\begin{equation}
 \Gamma^{(4)} = \frac{\delta^4 \Gamma_{int}}{\delta \phi^4} \Bigg|_{\bar{G}} +  
\frac{i}{2} \left[ \Lambda \bar{G}^2 \Lambda^{\dagger} + \frac{i}{2} \Lambda \bar{G}^2 \bar{V} \bar{G}^2 \Lambda^{\dagger} + perm  \right]. 
\label{4-pt-matr}
\end{equation}

In the exact theory, the following relation is satisfied:
\begin{equation} 
 \frac{\delta^2 \Gamma_{int}}{\delta \phi_1 \delta \phi_2} \Bigg|_{\phi_0 = 0} = 2 \, \frac{\delta \Gamma_{int}}{\delta G_{12}} \Bigg|_{\phi_0 = 0}
 \label{2-pt-relation2}
\end{equation}
which is the first consistency relation \eqref{2-pt-relation} we used. Differentiating with respect to $G$,
we obtain
\begin{equation}
 \frac{\delta^3 \Gamma_{int}}{\delta \phi_1 \delta \phi_2 \delta G_{34}} \Bigg|_{\phi_0 = 0} = 2 \, \frac{\delta^2 \Gamma_{int}}{\delta G_{12} \delta G_{34}} \Bigg|_{\phi_0 = 0}, 
\end{equation}
that is
\begin{equation}
 \Lambda_{12,34} = \bar{\Lambda}_{12,34}.
\label{lambdas}
\end{equation}
Given this, the 2-point function \eqref{2-pt} results
\begin{equation}
 \Gamma^{(2)}_{12} = i G_{0,12}^{-1} + 2 \frac{\delta \Gamma_{int}}{\delta G_{12}} \Bigg|_{\phi_0 = 0} = i \bar{G}^{-1}_{12},
\end{equation}
where the gap equation \eqref{gaux} was used to arrive at the last equality.

In a similar way, but now for  the  4-point function \eqref{4-pt-matr}, using \eqref{lambdas} and \eqref{Vert} we obtain
\begin{eqnarray}
 \Gamma^{(4)} &=& \frac{\delta^4 \Gamma_{int}}{\delta \phi^4} \Bigg|_{\bar{G}} +  
\frac{i}{2} \left[ \bar{\Lambda} \bar{G}^2 \bar{\Lambda}^{\dagger} + \frac{i}{2} \bar{\Lambda} \bar{G}^2 \bar{V} \bar{G}^2 \bar{\Lambda}^{\dagger} + perm  \right] \notag \\
&=& \frac{\delta^4 \Gamma_{int}}{\delta \phi^4} \Bigg|_{\bar{G}} + \frac{i}{2} \bar{V} \bar{G}^2 \bar{\Lambda}^{\dagger} + perm \\
&=& \frac{\delta^4 \Gamma_{int}}{\delta \phi^4} \Bigg|_{\bar{G}} + \left[ \bar{V} - \bar{\Lambda} + perm \right]
\end{eqnarray}
or, more explicitly
\begin{equation}
\Gamma^{(4)}_{1234} = \frac{\delta^4 \Gamma_{int}}{\delta \phi_1 \delta \phi_2 \delta \phi_3 \delta \phi_4} \Bigg|_{\bar{G}} + \left[ \bar{V}_{12,34} - \bar{\Lambda}_{12,34} + \bar{V}_{13,24} - \bar{\Lambda}_{13,24} + \bar{V}_{14,23} - \bar{\Lambda}_{14,23} \right] .
 \label{4-pt-relation2}
\end{equation}
This last equation is the second consistency condition we were looking for. In order to implement this, it is necessary to
consider another  relation which is valid for the exact theory:
\begin{equation}
  \Gamma^{(4)}_{12,34}(\phi_0=0) = \bar{V}_{12,34}(\phi_0=0),
\label{Vbar-Lambda}
\end{equation}
then the previous expression can be rewritten as
\begin{eqnarray}
\Gamma^{(4)}_{1234} = 2 \left[ \frac{\delta^2 \Gamma_{int}}{\delta G_{12} \delta G_{34}} \Bigg|_{\bar{G},\phi_0=0} + perms(2,3,4) \right] - \frac{1}{2} \frac{\delta^4 \Gamma_{int}}{\delta \phi_1 \delta \phi_2 \delta \phi_3 \delta \phi_4} \Bigg|_{\bar{G},\phi_0=0}, \,\,\,\,\,\,
 \label{4-pt-relation3}
\end{eqnarray}
where the symmetry properties of $\Gamma^{(4)}_{1234}$ where used. This is the second consistency relation \eqref{4-pt-relation} we used.

\section*{Appendix B: Relation among the renormalized and MS finite parameters}\label{MSvsRen}

In this section we summarize the calculation of the renormalized parameters defined from the effective potential, and we compute their relation with the MS  finite parameters.

For $m_R^2$ we use \eqref{ren-mass-def} together with \eqref{phys-mass-fin} and \eqref{F-prop1}, giving
\begin{equation}
 m_R^2 = m^2 + \frac{\lambda}{32 \pi^2} m_R^2 \, \ln \left( \frac{m_R^2}{\tilde{\mu}^2} \right)
\end{equation}
or, equivalently
\begin{equation}
 m_R^2 = \frac{m^2}{\left[ 1 - \frac{\lambda}{32 \pi^2} \ln \left( \frac{m_R^2}{\tilde{\mu}^2} \right) \right]}.
 \label{mR-mu2}
\end{equation}
Next, for $\xi_R$, we see from \eqref{ren-xi-def} that we must impose the condition $d m_{ph}^2/d R |_0 = 0$. For that we take the $R$-derivative of \eqref{ren-mass-def}
\begin{eqnarray}
\frac{dm_{ph}^2}{dR} + \xi_R& =& \xi + \frac{\lambda}{32\pi^2} \left(\xi_R - \frac{1}{6} \right) + \frac{\lambda}{32\pi^2} \left\{ \left[ \frac{dm_{ph}^2}{dR} + \xi_R - \frac{1}{6} \right] \ln \left( \frac{m_{ph}^2}{\tilde{\mu}^2} \right)\right. \nonumber\\
&+& \left.\left[ m_{ph}^2 + \left( \xi_R -\frac{1}{6} \right) R \right] \frac{1}{m_{ph}^2} \frac{dm_{ph}^2}{dR} - 2 \frac{dF}{dR}\right\}.
\end{eqnarray}
We can now evaluate for $\phi_0 = 0$ and $R=0$, use \eqref{ren-mass-def} and \eqref{F-prop3}, and impose the condition previously mentioned. This leads to
\begin{equation}
\xi_R = \xi + \frac{\lambda}{32\pi^2} \left(\xi_R - \frac{1}{6} \right) + \frac{\lambda}{32\pi^2}  \left( \xi_R - \frac{1}{6} \right) \ln \left( \frac{m_R^2}{\tilde{\mu}^2} \right) 
\end{equation}
or, with some algebra
\begin{equation}
\left( \xi_R - \frac{1}{6} \right) = \frac{\left( \xi - \frac{1}{6} \right)  }{ \left[1 - \frac{\lambda}{32\pi^2} - \frac{\lambda}{32\pi^2} \ln \left( \frac{m_R^2}{\tilde{\mu}^2} \right) \right] }.
\label{xiR-mu2}
\end{equation}
Finally for $\lambda_R$ we must take two $\phi_0$-derivatives of \eqref{phys-mass-fin},
\begin{eqnarray}
 \frac{dm_{ph}^2}{d\phi_0} &=& \lambda \phi_0 + \frac{\lambda}{32\pi^2} \left\{ \ln \left( \frac{m_{ph}^2}{\tilde{\mu}^2} \right) + \left[ m_{ph}^2 + \left(\xi_R - \frac{1}{6} \right) R \right] \frac{1}{m_{ph}^2}\nonumber\right.\\
 &-& \left. 2 \frac{dF(m_{ph}^2,\{R\})}{d m_{ph}^2} \right\} \frac{dm_{ph}^2}{d\phi_0}.
\end{eqnarray}
Before taking the second derivative, we evaluate this expression for $\phi_0 = 0$ and $R=0$ as it will be needed later
\begin{equation}
 \frac{dm_{ph}^2}{d\phi_0} \Bigg|_0 \left[ 1 - \frac{\lambda}{32\pi^2} - \frac{\lambda}{32\pi^2} \ln \left( \frac{m_{R}^2}{\tilde{\mu}^2} \right) \right] = 0
\end{equation}
which implies $\frac{dm_{ph}^2}{d\phi_0}|_0 = 0$. Now, deriving once more
\begin{eqnarray}
 \frac{d^2 m_{ph}^2}{d\phi_0^2} &=& \lambda + \frac{\lambda}{32\pi^2} \left\{ \ln \left( \frac{m_{ph}^2}{\tilde{\mu}^2} \right) + \left[ m_{ph}^2 + \left(\xi_R - \frac{1}{6} \right) R \right] \frac{1}{m_{ph}^2} \right.\nonumber\\
 &-&\left.  2 \frac{dF(m_{ph}^2,\{R\})}{d m_{ph}^2} \right\} \frac{d^2 m_{ph}^2}{d\phi_0^2} + \left[ \dots \right] \frac{d m_{ph}^2}{d\phi_0},
\end{eqnarray}
then evaluating for $\phi_0 = 0$ and $R=0$, using the previous result and \eqref{ren-lambda-def} we arrive at 
\begin{equation}
 \lambda_R = \frac{\lambda}{\left[ 1 - \frac{\lambda}{32\pi^2} - \frac{\lambda}{32\pi^2} \ln \left( \frac{m_{R}^2}{\tilde{\mu}^2} \right) \right]}.
\end{equation}

\section*{Appendix C: Coincident limit of the de Sitter propagator}\label{comparison}

Here we provide some details of the calculation of the function $F(m_{ph}^2,\{R\})$ for de Sitter spacetime. We must expand the de Sitter coincident propagator for $\epsilon \to 0$ and fixed $R$, and for this reason we rewrite $\nu_n^2 = \frac{(n-1)^2}{4} - \frac{m_{ph}^2}{H^2} - \xi_R n(n-1)$ as $\tilde{\nu}_n^2 = \frac{(n-1)^2}{4} - \left(\frac{m_{ph}^2}{R} + \xi_R \right) n(n-1)$. Then expanding for $\epsilon \to 0$:
\begin{eqnarray}
 [G_1] &=& \frac{R}{96 \pi^2} \frac{\Gamma \left( \frac{3}{2} + \nu_4 \right) \, \Gamma \left( \frac{3}{2} - \nu_4 \right)}{\Gamma \left( \frac{1}{2} + \nu_4 \right) \, \Gamma \left( \frac{1}{2} - \nu_4 \right)}  \Biggl\{ \frac{2}{\epsilon} - \frac{13}{6} + \gamma_E + \ln \left( \frac{R}{48\pi \mu^2} \right) + \psi\left( \frac{3}{2} + \nu_4 \right)  + \psi\left( \frac{3}{2} - \nu_4 \right)  \nonumber\\
 &+& 2 \left[ \psi\left( \frac{1}{2} - \nu_4 \right) - \psi\left( \frac{1}{2} + \nu_4 \right) + \psi\left( \frac{3}{2} + \nu_4 \right) - \psi\left( \frac{3}{2} - \nu_4 \right) \right] \frac{d\tilde{\nu}_n}{dn}\Bigg|_{n=4} \Biggr\} + \mathcal{O}(\epsilon) 
\end{eqnarray}
Here $\psi(x) = \Gamma^{'}(x)/\Gamma(x)$ is the DiGamma function, $\mu$ is an arbitrary mass scale introduced to maintain the usual units as $n\neq4$. Furthermore, although $\nu_4 = \tilde{\nu}_4$, the derivatives $d\nu_4/dn|_{n=4} \neq d\tilde{\nu}_4/dn|_{n=4}$. From now on $\epsilon$ can be set to $0$ in the non-divergent terms. Using the properties of the Gamma and DiGamma functions we can simplify the pre-factor and the expression between square brackets, leading to
\begin{equation}
 [G_1] = \frac{R}{96 \pi^2} \Biggl\{ \left( \frac{1}{4} - \nu_4^2 \right) \left[ \frac{2}{\epsilon} - \frac{13}{6} + \gamma_E + \ln \left( \frac{R}{48\pi \mu^2} \right) + \psi_{+} + \psi_{-} \right] - 4 \nu_4 \frac{d\tilde{\nu}_n}{dn}\Bigg|_{n=4} \Biggr\}
\end{equation}
where we use the condensed notation $\psi_{\pm} \equiv \psi\left( \frac{3}{2} \pm \nu_4 \right)$. Now replacing for $\nu_4$ and $d\tilde{\nu}_n/dn|_{n=4}$ we obtain
\begin{eqnarray}
 [G_1] &=& \frac{1}{8 \pi^2} \Biggl\{ \left[ m_{ph}^2 + \left( \xi_R - \frac{1}{6} \right) R \right] \left[ \frac{2}{\epsilon} - \frac{13}{6} + \gamma_E + \ln \left( \frac{R}{48\pi \mu^2} \right) + \psi_{+} + \psi_{-} \right] \nonumber\\
& -& \frac{R}{4} + \frac{7}{6} \left( m_{ph}^2 + \xi_R R \right) \Biggr\} 
\end{eqnarray}
or, rearranging slightly and substituting $\gamma_E - 1 + \ln(R/48\pi\mu^2) \to \ln(R/12\tilde{\mu}^2)$,
\begin{eqnarray}
 [G_1] &=& \frac{1}{4 \pi^2 \epsilon} \left[ m_{ph}^2 + \left( \xi_R - \frac{1}{6} \right) R \right] + \frac{1}{8 \pi^2} \Biggl\{ \left[ m_{ph}^2 + \left( \xi_R - \frac{1}{6} \right) R \right] \nonumber\\
 &\times&\left[ \ln \left( \frac{R}{12 \tilde{\mu}^2} \right) + \psi_{+} + \psi_{-} \right] - \frac{R}{18} \Biggr\}
\end{eqnarray}
In this form this expression is directly comparable to the first line of Eq. \eqref{adiab-G1}. The function $F(m_{ph}^2,\{R\})$ for de Sitter spacetime,
\begin{eqnarray}
F_{dS}(m_{ph}^2,R) &=& -\frac{1}{2} \left[ m_{ph}^2 + \left( \xi_R - \frac{1}{6} \right) R \right] \left[ \ln \left( \frac{R}{12 m_{ph}^2} \right) + \psi_{+} + \psi_{-} \right] \nonumber\\
&+& \frac{1}{2} \left( \xi_R - \frac{1}{6} \right) R + \frac{R}{36} 
\end{eqnarray}
where $R=12 H^2$. This function has all the expected properties, that is, it is written only in terms of renormalized parameters, it is independent of $\epsilon$ and $\tilde{\mu}$, and it satisfies the correct limits \eqref{F-prop1}, \eqref{F-prop2} and \eqref{F-prop3}. To verify these limits it is useful to consider that
\begin{subequations}
\begin{align}
\lim_{R \to 0} \left[ \psi_{+} + \psi_{-} + \ln \left( \frac{R}{12 m_{ph}^2} \right) \right] &= 0 \\
\lim_{R \to 0} \left[ m_{ph}^2 + \left( \xi_R - \frac{1}{6} \right) R \right] \left[ \frac{1}{R} + \left( \psi_{+}^{'} - \psi_{-}^{'} \right) \frac{d\tilde{\nu}_4}{dR} \right] &= \left( \xi_R - \frac{1}{6} \right) + \frac{1}{18}
\end{align}
\end{subequations}

\end{document}